\documentclass[12pt,amsfonts]{article}  %Moscow lecture material, August 2012
\usepackage{amsfonts}
\usepackage{amsmath}
\def\ts{\textstyle}
\def\t{\textstyle}        %DEAR EDITOR:  I PREFER A LARGER, TEXT SIZE FOR `STUFF' IN   $e^{STUFF}$  BUT IF

\def\one{1\hskip-.37em 1}

\def\half{{\textstyle{\frac{1}{2}}}}
\def\threebytwo{\textstyle{\frac{3}{2}}}
\def\quarter{\textstyle{\frac{1}{4}}}

\def\H{{\cal H}}

\def\threebytwo{\textstyle{\frac{3}{2}}}

\def\p{\phi}

\def\H{{\cal H}}
\def\hg{{\hat g}}
\def\hp{{\hat\pi}}
\def\hph{{\hat\phi}}

\def\g{\gamma}

\def\l{\lambda}
\def\D{{\cal D}}

\def\S{\Sigma'}

\def\op{{\overrightarrow p}}
\def\oq{{\overrightarrow q}}
\def\oP{{\overrightarrow P}}
\def\oQ{{\overrightarrow Q}}
\def\ox{{\overrightarrow x}}
\def\oy{{\overrightarrow y}}
\def\oR{{\overrightarrow R}}
\def\oS{{\overrightarrow S}}

\def\k{\kappa}

\def\bx{x}   %\def\bx{{\bf x}}

\def\E{{\rm I}\hskip-.2em{\rm E}}
\def\ra{\rightarrow}
\def\threequarters{{\textstyle{\frac{3}{4}}}}

\def\hr{\hat{\rho}}
\def\tint{{\textstyle\int}}
\def\hg{{\hat g}}
\def\hp{{\hat\pi}}

\def\hph{{\hat\phi}}
\def\s{\hskip.08em}
\def\d{\partial}
\def\o{\overline}
\def\a{\alpha}
\def\b{\begin{eqnarray}}  %takes no eqn numbers
\def\e{\end{eqnarray}}    %takes no eqn numbers
\def\bn{\begin{eqnarray}}  %takes eqn numbers
\def\en{\end{eqnarray}}   %takes eqn numbers
\def\<{\langle}
\def\>{\rangle}

\def\no{\nonumber}

\def\k{\kappa}

\def\de{\delta}

\def\{{\lbrace}
\def\}{\rbrace}
\title{Enhanced Quantum Procedures that \\Resolve Difficult Problems\footnote{Based on a lecture for the Advanced Scholar Seminar, ``Interaction of Mathematics and Physics: New Perspectives'', Moscow, Russia, August, 2012. The lecture may be viewed at $http://www.mathnet.ru/php/conference.phtml?\& eventID=1\& confid=297\& option\_lang=eng$ (lecture 18).
}}
\author{John R. Klauder\footnote{Email: klauder@phys.ufl.edu}\\
Department of Physics and\\Department of Mathematics\\
University of Florida\\
Gainesville, FL 32611-8440
 }
\date{ }
\numberwithin{equation}{section}          %for section numbering of equations
\begin{document}
\maketitle
\begin{abstract}
A careful study of the classical/quantum connection with the aid of coherent states offers new insights into various technical problems. This analysis includes both canonical as well as closely related affine quantization procedures. The new tools are applied to several examples including:
  (1) A quantum formulation that is invariant under arbitrary classical canonical transformations of coordinates;
  (2) A toy model that for all positive energy solutions has singularities which are removed at the classical level when the correct quantum corrections are applied;
  (3) A fairly simple model field theory with nontrivial classical behavior that, when conventionally quantized, becomes trivial, but nevertheless  finds a proper solution using the enhanced procedures;
  (4) A model of scalar field theories with nontrivial classical behavior that, when conventionally quantized, becomes trivial, but nevertheless  finds a proper solution using the enhanced procedures;
  (5) A viable formulation of the kinematics of quantum gravity that respects the strict positivity of the spatial metric in both its classical and quantum versions; and
  (6) A proposal for a nontrivial quantization of $\phi^4_4$ that is ripe for study by Monte Carlo computational methods.
    All of these examples use fairly general arguments that can be understood by a broad audience.
   \end{abstract}
   \tableofcontents
 \section{An Overview of \\Canonical and Affine Quantization}
 From a conceptual as well as an operational viewpoint classical mechanics (also classical field theory) and quantum mechanics (also
 quantum field theory) are quite well established and automatically brought to bear on any new problem. In large measure, the successes of the canonical view for classical and quantum theories are universal, and when the canonical procedures seem to fail, it is not the canonical procedures themselves that are questioned but rather the suitability of the problem under consideration. Such problems include---but are not limited to---situations in which a nonlinear and nontrivial classical model is rendered trivial (equivalent to a free model) upon quantization. Or models that have well behaved classical behavior but exhibit perturbation series in their quantum formulation that entail infinitely many distinct divergences. Such misbehaving models are shunned for their
 nonconformity and relegated to the waste basket in hope that they may never be encountered in the real world. But
 the purest form of quantum gravity is one of these misbehaving theories and its difficulties upon quantization have led others to require %extraordinary
 a new and fundamentally different formulation  replete with exotic overlays of additional fields
  to overcome its bad behavior (superstring theory). Yet another quantum theory of gravity chooses unusual
 basic kinematical variables that seem remote from the usual classical variables themselves (loop quantum gravity). These statements are only meant to show the extent needed to obtain viable candidates

 Our goal in this presentation is to slowly develop an alternative prescription regarding the standard classical and quantum connection as well as the quantization procedure itself which demonstrate that several difficult problems as well as the truly difficult problem of quantizing the gravitational field of Einstein can be carried out without the need for any exotic features whatsoever.

 However, before we can discuss the most difficult problems we need to build a philosophy that makes the alternative prescription of the standard procedures seem almost inevitable. This we do by introducing the new aspects in several simple models that will develop the new connection procedures as we go along. To begin this adventure into {\it terra incognita}, we need to establish our own approach toward classical and quantum theories. The formulation we shall develop is by now no longer new but has been developing slowly for several decades. In the first section of this article, we shall illustrate the needed formulation in a simple but general enough fashion so that it will be more natural to analyze the variety of problems that we wish to entertain. The presentation in Sec.~1 has been inspired by \cite{CRTII,CRTIII,acsIOP,MOSCOW}.

 \subsection{Canonical quantization}
 The standard way to start the quantization of a classical system is to first agree on a suitable set of kinematical variables. In the Hamiltonian formulation of classical mechanics for a single degree of freedom, one introduces the classical
 momentum $p$ and the classical position $q$. We initially focus on what we call Cartesian coordinates (more on this subject later). Classically, we introduce the concept of a Poisson bracket, denoted by $\{\;,\;\}$, that operates on two functions of $p$ and $q$, say $A=A(p,q)$ and $B=B(p,q)$, and which is then defined by
   \bn \{\s A\s,\s B\s\}\equiv \frac{\d A}{\d q}\,\frac{\d B}{\d p}-\frac{\d A}{\d p}\,\frac{\d B}{\d q}\;. \en
   It follows that the Poisson bracket $\{\s q\s,\s p\s\}=1$, and any two variables that have a Poisson bracket
   that is unity are called canonically conjugate variables. One such example is given by ${\tilde p}=p\s q^{-2}/3$ and ${\tilde q}=q^3$, since it follows that
       \bn \{\s{\tilde q}\s,\s{\tilde p}\}= \frac{\d \s (q^3)}{\d q}\,\frac{\d\s (p\s q^{-2}/3)}{\d p}-\frac{\d \s (q^3)}{\d p}\,\frac{\d\s (p\s q^{-2}/3)}{\d  q}=1\;,\en
   as required. The full space covered by the coordinates $p$ and $q$ is called phase space and these coordinates are
   called phase space or canonical variables. A scalar on phase space is a function, say $A(p,q)$, that transforms as a
   scalar, namely, under a canonical transformation from one set of canonical coordinates to another set of canonical coordinates, such as $(p,q)\ra({\tilde p},{\tilde q})$, then ${\tilde A}({\tilde p},{\tilde q})=A(p,q)$. An abstract setting may also be introduced. Let ${\cal M}$ denote a two-dimensional topological
   space, endowed with a concept of continuity but without a metric structure. We impose one set of canonical coordinates $(p,q)$ on ${\cal M}$ to determine a point ${\cal P}\in{\cal M}$, and we also choose another set of canonical coordinates $({\tilde p},{\tilde q})$ that determine the {\it same} point ${\cal P}\in {\cal M}$.
   Additionally, there is an abstract scalar ${\cal A}$ on ${\cal M}$ that has a specific value at the point ${\cal P}$, and which we identify either by $A(p,q)$ or by ${\tilde A}({\tilde p},{\tilde q})$. Although there is
   an abstract, coordinate-free formulation of these matters, it is traditional to adopt an appropriate set of canonical coordinates for computational purposes, as we shall do hereafter.

   In suitable canonical coordinates (defined as Cartesian) the canonical quantization procedure {\it promotes}, the variables $p$ and $q$ to operators $P$ and $Q$ acting in a Hilbert space ${\frak H}$.
   These operators are minimally required to be Hermitian (symmetric in the mathematics literature) and to obey the Heisenberg commutation relation $[\s Q\s, P\s]\equiv Q\s P-P\s Q =i\s\hbar\s\one$, where  $\hbar\simeq 10^{-27} erg\s\cdot\s sec$ is the reduced Planck constant and $\one$ is the unit operator. There are many common representations of $P$ and $Q$.
   For example, $Q\ra x$, $x\in{\mathbb R}$, and $P\ra -i\s\hbar\s\d/\d x$. This representation works for any interval such as  $-X<x<X'$. However, for these operators to play their most valuable role, they must be self adjoint, which is a stronger condition than being Hermitian, and to attain this criterion entails suitable restrictions on the boundary conditions. For the present, we shall be interested in the case that $-\infty<x<\infty$, for which there is only one irreducible representation (up to unitary equivalence) and it is the representation given above. In that case both $P$ and $Q$ are uniquely determined such that the operators $\exp(-iqP/\hbar)$ and $\exp(ip\s Q/\hbar)$ are both unitary for all $q$ and $p$, respectively; $U$ is unitary provided $U\s U^\dag=U^\dag\s U=\one$, where $U^\dag$ denotes the adjoint operator. Assuming that
     \bn U[p,q]\equiv e^{\t -iq P/\hbar}\s e^{\t ip\s Q/\hbar}= e^{\t -ip\s q/\hbar}\s e^{\t ip\s Q/\hbar}\s e^{\t -iqP/\hbar}\;,\en
we deal with basic kinematical operators $P$ and $Q$ each having the whole real line as their spectrum.
     Formal eigenvectors exist such that $Q\s|x\>=x\s|x\>$ and $P\s |p\>=p\s|p\>$ for which $\<x|x'\>=\delta(x-x')$,  $\<p|p'\>=\delta(p-p')$, and $\<x|p\>=\exp(ip\s x/\hbar)/\sqrt{2\pi\hbar}$. Additionally, $\one=\tint |x\>\<x|\s dx=\tint |p\>\<p|\s dp$ when integrated over the whole real line in each case.

     Canonical coherent states involve the unitary operator $U[p,q]$ acting on a normalized, fiducial Hilbert space vector $|\eta\>\in {\frak H}$ and are defined as
        \bn |p,q\>\equiv U[p,q]\s|\eta\>=e^{\t -iqP/\hbar}\,e^{\t ipQ/\hbar}\,|\eta\>\;, \en
        where the dependence on the fiducial vector is generally left implicit. The coherent states have several
        interesting properties that hold for a general unit vector $|\eta\>$; see, e.g., \cite{skag}. In particular, the
        canonical coherent states
        admit a resolution of unity of the form
          \bn \one=\tint |p,q\>\<p,q|\,dp\, dq/2\pi\hbar \en
          when integrated over the entire phase space ${\mathbb R}^2$.
          Moreover, the diagonal matrix elements such as
             \bn A(p,q)\equiv \<p,q|\s {\cal A}\s|p,q\>\;, \en
             uniquely determine the operator ${\cal A}$ provided the coherent states are in its domain and ${\cal A}$ is a polynomial in irreducible operators, $P$ and $Q$. If the coherent state overlap function $\<p,q|p',q'\>$ never vanishes, then the diagonal elements determine the operator uniquely provided the coherent states are in its domain and $P$ and $Q$ are irreducible. As another interesting property, we note that
              \bn {\cal B}\equiv \int b(p,q)\,|p,q\>\<p,q|\,dp\, dq/2\pi\hbar  \en
              is a valid and unique representation of the operator ${\cal B}$ provided the coherent states are in its domain and the operator ${\cal B}$ is a polynomial in irreducible operators, $P$ and $Q$. Again, if the coherent state overlap function nowhere vanishes, then every operator in the domain of the coherent states can
              be represented in this fashion.  It is common to choose the fiducial vector $|\eta\>=|0\>$ as the ground state of a special harmonic oscillator such that $(Q+i\s P)\s|0\>=0$. In this case the coherent state overlap function is given by
                \bn \<p,q|p',q'\>=\exp\{(i/2\hbar)(p+p')(q-q')-(1/4\hbar)[(p-p')^2+(q-q')^2]\}\;, \en
                which vanishes nowhere. A weaker requirement, which is nevertheless still useful, is for $P$ and $Q$ to be ``physically centered'', which means that $\<\eta|\s P\s|\eta\>=\<\eta|\s Q\s|\eta\>=0$, and in this case, $\<p,q|\s P\s|p,q\>=p$ and $\<p,q|\s Q\s|p,q\>=q$. In what
                follows we will mostly focus on choosing $|0\>$ as the fiducial vector.

                So far we have concentrated on kinematics, which, of course, is an important part of the story. Now it is time to take up dynamics. Let us consider the classical mechanics of a single degree of freedom. We introduce a scalar function $H_c(p,q)$ which is called the Hamiltonian. It is the generator of motion in time, and  we adopt the standard equations of motion
                 \bn {\dot q}\hskip-1.4em&&=\frac{\d H_c(p,q)}{\d p}=\{\s q\s,\s H_c(p,q)\s\}\;,\no\\
                      {\dot p}\hskip-1.4em&&=-\frac{\d H_c(p,q)}{\d q}=\{\s p\s,\s H_c(p,q)\s\}\;,\en
                      where ${\dot q}\equiv d\s q/d\s t$, etc. If $H_c(p,q)$ is independent of the time $t$, then
          it follows that
        \bn {\dot H}_c(p,q)={\dot p}\,\frac{\d H_c(p,q)}{\d p}+{\dot q}\,\frac{\d H_c(p,q)}{\d q}={\dot p}\s{\dot q-{\dot q}{\dot p}}=0\;,\en
        and thus $H_c(p,q)=E$, where the energy $E$ is a constant of the motion. It is noteworthy that the Hamiltonian equations of motion are invariant under canonical transformations, and in particular
           \bn {\dot {\tilde q}}\hskip-1.4em&&=\frac{\d {\tilde H}_c({\tilde p},{\tilde q})}{\d {\tilde p}}=\{\s {\tilde q}\s,\s {\tilde H}_c({\tilde p},{\tilde q})\s\}\;,\no\\
                      {\dot{\tilde p}}\hskip-1.4em&&=-\frac{\d {\tilde H}_c({\tilde p},{\tilde q})}{\d {\tilde q}}=\{\s {\tilde p}\s,\s {\tilde H}_c({\tilde p},{\tilde q})\s\}\;.\label{cc}\en

                     Finally, we introduce a very important quantity, namely, a classical variational principle from which the equations of motion can be derived. Consider the action functional defined on phase space paths $(p(t),q(t))$ and given by
                       \bn A_C =\tint_0^T[p(t)\s{\dot q}(t)-H_c(p(t),q(t))]\,dt\;, \en
                       and consider its variation (the difference between infinitesimally nearby paths) which is given by
                       \bn  \delta A_C=\tint_0^T\{[{\dot q}-\d H_c(p,q)/\d p]\,\delta p +[-{\dot p}-
                       \d H_c(p,q)/\d q]\,\delta q\,\}\,dt+ p\,\delta q\big|^T_0\;.\en
                       We insist that this first-order variation vanish for arbitrary path variations that also satisfy
                       $\delta q(0)=\delta q(T)=0$, and for reasons that are not evident at this point, we also require that
                       $\delta p(0)=\delta p(T)=0$ as well. It follows that the condition $\delta A_C=0$,  leads to the classical equations of motion, and unique solutions to those equations of motion  are obtained, e.g.,  by choosing $(p(0),q(0))$ as initial data. We can also introduce new canonical coordinates as before, now using the relation
                       \bn p\s\s dq={\tilde p}\s\s d{\tilde q}+d{\tilde G}({\tilde p},{\tilde q}) \en
                       among one forms for some generator ${\tilde G}({\tilde p},{\tilde q})$. (For our specific example, where
                       ${\tilde p}=p\s q^{-2}/3$ and ${\tilde q}=q^3$, it follows that ${\tilde G}=0!$) With a general change of canonical variables in mind, the classical action becomes
                         \bn A_C=\tint_0^T[\s{\tilde p}\s{\dot{\tilde q}}+{\dot{\tilde G}}(\tilde p,{\tilde q})
                         -{\tilde H}_c({\tilde p},{\tilde q})\s]\,dt\;, \en
                         where ${\tilde H}_c({\tilde p},{\tilde q})\equiv H_c(p,q)$, and a stationary variation of this equation holding both ${\tilde p}$ and ${\tilde q}$ fixed
                         at $t=0$ and $t=T$, leads to the proper equations of motion (\ref{cc}) in the new coordinates
                         to which ${\tilde G}({\tilde p},{\tilde q})$ makes no contribution. Again, unique solutions to those equations of motion follow by choosing $({\tilde p}(0),{\tilde q}(0))$ as initial data.

                          Let us return to the quantum story in which we now add dynamics to our previous kinematical discussion. Associated with the classical Hamiltonian $H_c(p,q)$ is a quantum Hamiltonian operator $\H=\H(P,Q)$. If the right coordinates are chosen, the quantum Hamiltonian operator is given by $\H=H_c(P,Q)+\hbar\s Y(P,Q)$, where the last factor $Y$ accounts for possible factor ordering as well as other ambiguities that arise
                         in the very process of quantization. Normally, in the classical limit in which $\hbar\ra0$, the factor $Y$ generally makes no contribution to the classical limit and thus $Y$ may be open to different choices. Indeed, it may even be possible that different physical applications of the same classical model problem could have different choices for the auxiliary term $Y$ when quantized. Thus it is prudent to keep such
                         auxiliary terms in mind so that they may be exploited when necessary. For simplicity,  we shall presently assume that the quantum Hamiltonian is given by $\H=H_c(P,Q)$ (modulo factor ordering if needed to ensure an Hermitian operator) which is true for many systems.  The normalized state vector $|\psi(0)\>\in{\frak{H}}$ is assumed to represent  the state of the system at the initial time $t=0$, and we are interested in identifying the state vector $|\psi(t)\>$ that represents the state of the system at a future time $t>0$. While the Hamiltonian operator $\H$ may be Hermitian, it is important to ensure that it becomes self adjoint and thus has a real spectrum and can serve as the generator of a unitary operator such as $U(t)=\exp(-i\H\s t/\hbar)$ for all real $t$. Thus, the abstract evolution equation takes the form
                         \bn |\psi(t)\>=e^{\t -i\H\s t/\hbar}\,|\psi(0)\>\;, \en
                         and the abstract Schr\"odinger's equation is given by the time derivative of this equation, namely
                           \bn i\s\hbar\, \d \s|\psi(t)\>/\d t=\H\,|\psi(t)\> \;.\en
                         Various representations of this equation can be introduced by using the states
                         $|x\>$, $|p\>$, and $|p,q\>$. Thus, assuming $\H=H_c(P,Q)$, we are led to
                           \bn i\s\hbar\s\s\d\s\psi(x,t)/\d t\hskip-1.4em&&=H_c(-i\hbar\s \d/\d x,x)\,\psi(x,t)\;, \no\\
                              i\s\hbar\s\s\d\s\psi(p,t)/\d t\hskip-1.4em&&=H_c(p,i\s\hbar \d/\d p)\,\psi(p,t)\;, \\
                              i\s\hbar\s\s\d\s\psi(p,q,t)/\d t\hskip-1.4em&&=H_c(-i\s\hbar \d/\d q,q+i\s\hbar \d/\d p)\,\psi(p,q,t)
                              \;,\no\en
                              where $\psi(x)\equiv \<x|\psi\>$, $\psi(p)\equiv\<p|\psi\>$, and $\psi(p,q)\equiv\<p,q|\psi\>$. The physical meaning of the various coordinates can be different in the different representations; namely, since $Q\s|x\>=x\s|x\>$ and $P\s|p\>=p\s|p\>$, it follows that $x$ and $p$ refer to {\it sharp eigenvalues}. On the other hand, and assuming that $|\eta\>$ is physically centered,
                              %so that $\<\eta|P|\eta\>=\<\eta|Q|\eta\>=0$, we find
                              it follows that $\<p,q|\s P\s|p,q\>=p$ and $\<p,q|\s Q\s|p,q\>=q$, and thus in the coherent state representation, $p$ and $q$ are {\it mean values}, and
                              thus they can be simultaneously specified.

                              It is noteworthy that we can derive the abstract quantum equation of motion from a variational principle. Specifically,
                              the quantum action functional is given by
                              \bn A_Q=\tint_0^T\,\<\psi(t)|\s[\s i\s\hbar\s\d/\d t-\H(P,Q)\s]\s|\psi(t)\>\,dt\;,\en
                              and we may confine attention to unit vectors $|\psi(t)\>$.
                              Treating $\<\psi|$ and $|\psi\>$ as independent variables, a first-order variation leads to
                              \bn &&\delta A_Q=\tint_0^T\,\{\,(\delta\<\psi|)[i\hbar\overrightarrow{\d} /\d t-\H]|\psi\>+
                                \<\psi|[-i\hbar\overleftarrow{\d}/\d t-\H](\delta|\psi\>)\,\}\,dt\no\\
                                &&\hskip9em +i\hbar\<\psi|\,\delta|\psi\>\big|^T_0\;. \en
                    Insisting that this expression vanish for general variations $\delta\<\psi(t)|$ and $\delta|\psi(t)\>$, while requiring that $\delta|\psi(0)\>=\delta|\psi(T)\>=0$, leads to the abstract Schr\"odinger equation
      \bn i\hbar\s\s\s\d \s|\psi(t)\>/\d t=\H\,|\psi(t)\>\en
      as well as its adjoint. Choosing $|\psi(0)\>$, or its adjoint, as initial data leads to a unique solutions to those equations of motion.

\subsubsection{Coexistence of classical and quantum mechanics}
      In typical classical/quantum stories, one has to take the limit $\hbar\ra0$  to recover the classical theory from the quantum theory. But in fact, in the real world, $\hbar>0$  and we need an approach in which {\it both the classical theory and the quantum theory coexist at the same time}. We assert that
      we can achieve this desired feature in the following way. Rather than allowing an {\it arbitrary} set of quantum states  in the quantum action functional, we {\it restrict} the set of allowed quantum states to those that can be macroscopically created, namely, those that translate a system or put it into motion with a constant velocity [the latter based on the fact that ${\dot q}=\d H_c(p,q)/\d p\s$]. In other words, we restrict the allowed states in the quantum action principle so that
        \bn |\psi(t)\>\ra |p(t),q(t)\>\;, \en
        that is, {\it just to the coherent states}. This leads to the {\it restricted quantum action functional} given by
          \bn A_{Q(R)}=\tint_0^T\,\<p(t),q(t)|[\s i\hbar \d/\d t-\H\s]|p(t),q(t)\>\,dt\;.\en
          The first term is
          \bn \<p(t),q(t)|[\s i\hbar \d/\d t]|p(t),q(t)\>\hskip-1.3em&&=\<\eta|\s {\dot q}\s(P+p)\s|\eta\>-\<\eta|\s Q\s|\eta\>{\dot p}\no\\
          &&=p(t)\s {\dot q}(t)+ \<\eta|\s P\s|\eta\>{\dot q}-\<\eta|\s Q\s|\eta\>{\dot p}\no\\
           &&=p\s{\dot q} \;,\en
          ignoring the total derivatives or adopting a physically centered fiducial vector, and the second term leads to
          \bn H(p(t),q(t))\equiv \<p(t),q(t)|\H|p(t),q(t)\>\;, \en
          which yields a real expression that has the appearance of the classical Hamiltonian, but there is  the
          very important difference that $\hbar$ is {\it not zero}. Thus the expression for the restricted quantum action functional (omitting the total derivatives $\<\eta|\s P\s|\eta\>\,{\dot q}$ and  $-\<\eta|\s Q\s|\eta\>{\dot p}$) becomes
            \bn A_{Q(R)}=\tint_0^T[\s p(t)\s\s {\dot q}(t)-H(p(t),q(t))\s]\,dt\;,\en
            which has every right to be called the classical action functional,  or better  {\it the enhanced classical action functional} reflecting the fact that $\hbar>0$. For an irreducible representation of $P$ and $Q$, the enhanced classical Hamiltonian becomes
            \bn H(p,q)\hskip-1.4em&&=\<p,q|\s\H(P,Q)\s|p,q\>\no\\
            && =\<\eta|\s\H(P+p,Q+q)\s|\eta\>\no\\
            &&=\H(p,q)+{\cal O}(\hbar;p,q)\;,\en
            a relation that is one part of the {\it Weak Correspondence Principle} \cite{WCP}.
       Here, in the last relation, we have assumed that $\<\eta|(P^2+Q^2)|\eta\>\ra0$ as $\hbar\ra0$, and that the coherent states are in the domain of the Hamiltonian operator $\H$. Thus, in these particular phase-space coordinates, the leading term (${\cal O}(\hbar^0$)) of the enhanced classical Hamiltonian is exactly the quantum Hamiltonian evaluated at the $c$-numbers $p$ and $q$. It is these particular coordinates that are deemed to be ``Cartesian'',
       and we are now in a position to clarify that remark. We have already used the fact
       that the one form $i\hbar\<p,q|\,d|p,q\>=p\s\s\s dq$ when $|\eta\>$ is physically centered, and we now investigate the metric
       that arises from the coherent states themselves. In particular, assuming the necessary domain conditions and using $\<(\cdot)\>=\<\eta|(\cdot)|\eta\>$ for convenience, the coherent-state-induced, Fubini-Study metric (which vanishes for changes only by a phase) becomes
        \bn d\sigma^2(p,q)\hskip-1.4em&&\equiv (2\s\hbar)[\s \|\s d|p,q\>\s\|^2-|\<p,q|\s\s d|p,q\>|^2\s]\no\\
           &&=(2/\hbar)[ \<Q^2\>\,dp^2+\<(Q\s P+ P\s Q)\>\,dp\s\s dq+\<P^2\>\,dq^2]\;, \en
            corresponding to a {\it flat} two-dimensional space. For the common choice of
           $|\eta\>=|0\>$, the metric becomes simply $d\sigma^2(p,q)=dp^2+dq^2$, which is strictly Cartesian in form.
If one changed from these canonical coordinates to other canonical coordinates, then the one form would possibly develop an exact differential, while the metric would still describe a flat space, but, generally speaking, it would be expressed in curvilinear coordinates and no longer in Cartesian
coordinates. While this metric has arisen from the Hilbert space geometry of the coherent state rays, this metric may be carried over to the classical phase space, and it is in this sense that the present classical coordinates, $(p,q)$, are Cartesian.

\subsubsection{General canonical coordinates}
Of course, we can also describe the enhanced classical theory in arbitrary canonical coordinates as follows. Let both $(p,q)$ and $({\tilde p},{\tilde q})$ be a pair of labels, from different canonical coordinate systems, for the same point ${\cal P}$ in the phase space ${\cal M}$. We purposely define the coherent states in the coordinates $({\tilde p},{\tilde q})$ by
\bn |{\tilde p},{\tilde q}\>\equiv|p({\tilde p},{\tilde q}),q({\tilde p},{\tilde q})\>=|p,q\>\;,\en
i.e., we define the coherent states as {\it scalars under canonical coordinate transformations}. Thus, it follows that
  \bn A_{Q(R)}\hskip-1.4em&& =\tint_0^T\<{\tilde p}(t),{\tilde q}(t)|[\s i\hbar\s\d/\d t-\H(P,Q)\s]|{\tilde p}(t),{\tilde q}(t)\>\,dt\no\\
    &&=\tint_0^T[\s {\tilde p}(t)\s{\dot{\tilde q}}(t)+{\dot{\tilde G}}({\tilde p}(t),{\tilde q}(t))-{\tilde H}({\tilde p}(t),{\tilde q}(t))\s]\,dt\;,\en
    where
    \bn {\tilde H}({\tilde p},{\tilde q})\hskip-1.4em&&=\<{\tilde p},{\tilde q}|\H(P,Q)|{\tilde p},{\tilde q}\>\no\\
    &&=\<\eta|\H(P+p({\tilde p},{\tilde q}),Q+q({\tilde p},{\tilde q}))|\eta\>\no\\
   &&=H(p({\tilde p},{\tilde q}),q({\tilde p},{\tilde q})) \;,\en
    showing that {\it the enhanced quantization procedure is completely invariant under canonical coordinate transformations}. Furthermore, {\it no} change of the operators arises in a canonical coordinate transformation.

    It is noteworthy, at this point, to observe that we have obtained basically the standard formulation of canonical quantization while clarifying the meaning of Cartesian coordinates  and improving the general story with regard to canonical transformations.

\subsubsection{Dilated fiducial vectors}
Although we have focussed largely on the fiducial vector $|0\>$ defined by $(Q+iP)\s|0\>=0$, we should also
say a few words about fiducial vectors $|0;\Omega\>$ that are defined by $(\Omega\s Q+iP)\s|0;\Omega\>=0$ for $\Omega\not=1$. The Fubini-Study metric in such cases would become $d\sigma^2(p,q)=\Omega^{-1}\s dp^2+\Omega\, dq^2$, which remains quasi-Cartesian. However, there is another feature that is important to mention. For traditional classical Hamiltonians of the form $H_c(p,q)=p^2+V(q)$ (for simplicity, choosing units so that the mass equals 1/2), it is frequently physically correct to choose the Hamiltonian operator to be $\H=P^2+V(Q)$. For such examples, the enhanced classical Hamiltonian becomes
    \bn H(p,q)\hskip-1.4em&&=\<p,q;\Omega|\s[\s P^2+V(Q)\s]\s|p,q;\Omega\>\no\\
    &&=p^2+\half\Omega\hbar+(\Omega/\pi\hbar)^{1/2}\tint V(x+q)\,e^{-\Omega\s x^2/\hbar}\,dx\;, \en
    which, apart from a constant, can be brought arbitrarily close to $p^2 +V(q)$ as desired by
    choosing a sufficiently large value for $\Omega$.

\subsubsection{Introduction of the Lagrangian}
It is noteworthy that we can also recover the enhanced Lagrangian action functional for suitable systems from the restricted quantum action functional. The equation ${\dot q}\equiv \d H(p,q)/\d p\s$ can be used to eliminate the variable $p$
for a simple set of systems as follows. If the restricted action functional has the form $\tint[\s p\s{\dot q}-
p^2/2m-{\widetilde V}(q)\s]\,dt$, then we can use the equation of motion ${\dot q}=p/m$ to obtain the Lagrangian form of the action functional as $\int[\s\half\s m{\dot q}^2-{\widetilde V}(q)\s]\,dt$, which implies, for the restricted quantum action, that
  \bn A_{Q(RL)}\hskip-1.3em&&=\tint_0^T\<m{\dot q}(t),q(t)|[\s i\s\hbar\s\d/\d t-P^2/2m-V(Q)\s]|m{\dot q}(t),q(t)\>\,dt \no\\
     &&=\tint_0^T[\s\half\s m{\dot q}(t)^2-{\widetilde V}(q(t))\s]\,dt\;. \en
The variational principle for this action requires that $\delta q(0)=\delta q(T)=0$, but specifying
the values of $q(0)$ and $q(T)$ as initial data does {\it not}, in general, ensure a unique solution for $q(t)$, $0\le t\le T$; instead, one must choose, e.g., $({\dot q}(0),q(0))$ as initial data to ensure uniqueness. In this case,
     the Lagrangian form of the action leads  to the same variational coherent states that would be found from the Hamiltonian form of the action.

     We now return to the Hamiltonian formulation, and we do not discuss the Lagrangian
     formulation any further.

\subsection{Affine quantization}
     If we restrict attention to standard canonical quantization in which $p\ra P$ and $q\ra Q$ for which
     $[Q,P]=\i\hbar\one$, then the analysis given previously would be the whole story. However, as emphasized several times, the standard canonical quantization is not a procedure that permits the classical and quantum stories to coexist. Looked at from the enhanced quantization point of view,
      the two-dimensional sheet of Hilbert space vectors represented by the set of canonical coherent states $|p,q\>$ for all $(p,q)\in{\mathbb R}^2$ may not be the whole story encompassed by enhanced quantization. In fact, there is a very different two-dimensional sheet of Hilbert space vectors that is of considerable interest, and let us discuss how we may find this interesting, alternative sheet in  Hilbert space. As noted earlier, both of the operators $P$ and $Q$ must be self adjoint in order to serve as generators of unitary operators; being Hermitian is necessary, but is not sufficient to generate unitary transformations. However, self-adjoint operators $P$ and $Q$ are {\it not} possible if (say) $Q>0$,  corresponding to a classical limitation that  $q>0$.
     If that is the case, it is impossible to arrange that the conjugate operator $P$ is self adjoint, and this is because if $P=-i\hbar \d/\d x$, then $\exp(-x)$ is a normalizable eigenfunction with an imaginary eigenvalue. The same situation holds if $Q>-\gamma\s\one$; however, we will continue to focus on the case where $Q>0$. This situation raises the question of how to deal with those cases in which the spectrum of one canonical operator (chosen here as $Q$) is confined for physical reasons to the positive half line.

     Of course, we wish to depart from the canonical quantization rules as little as necessary, and, if possible, we  even wish  to comply with them. Thus we seek a modification of the quantization procedure that is in fact implied by canonical quantization itself. An initial remark about a modification of the classical kinematical variables consistent with the standard classical canonical variables is in order. The Poisson bracket for the canonical variables $q$ and $p$ is $1=\{q,p\}$. Let us multiply this relation by $q$ leading to $q=q\{q,p\}=\{q,p\s q\}$, i.e., $\{q,d\}=q$, where $d\equiv p\s q$, and $d$ and $q$ may well serve as an alternative pair of basic kinematical variables. Observe that $d$ serves to {\it dilate} $q$, rather than {\it translate} $q$ as $p$ does; dilation of $q$ is compatible with the restriction $q>0$. With this much motivation, we propose to consider this line of argument next in terms of quantum variables.

     Let us take the Heisenberg commutation relation $i\hbar\one=[Q,P]$ and multiply both sides by $Q$
     leading to $ i\hbar\s Q=Q\s[Q,P]=[Q,QP]=[Q,\half(PQ+QP)]$,
       or stated otherwise, we are led to
       \bn  [Q,D]=i\hbar\s Q\;,\hskip3em  D\equiv\half(PQ+QP)\;. \en
       The variables  $Q$ and $D$ are called {\it affine variables} and the Lie algebra commutation relation $[Q,D]=i\hbar\s Q$ is called an {\it affine commutation relation} since it refers to the affine group of transformations: $x\ra x'=a\s x+b$, where $a\not=0$ and $x,a,b\in{\mathbb R}$. There are three inequivalent, irreducible representations of the affine Lie algebra, one for $Q>0$, one for $Q<0$, and one where $Q=0$. For each of these representations, the operators $Q$ and $D$ can both be self adjoint, and we shall choose them to be self adjoint. Just as in the classical discussion above,
       the physics leading to favor $D$ over $P$ is that $D$ acts to dilate $Q$ while $P$ acts to  translate $Q$. There are occasions where it is necessary to consider all three inequivalent representations, but, for the present,
       our interest will be focussed on the choice $Q>0$. The affine commutation relation ensures that the dimensions of $D$ are those of $\hbar$, while the dimensions of $Q$ are open. Normally, we will be interested in $Q/q_0$  being dimensionless, for some positive reference value $q_0$, but for convenience
       we will generally work in units where $q_0=1$, With $-\infty<p<\infty$ and $0<q<\infty$, we introduce two unitary one-parameter families given by $\exp(ip\s\s Q/\hbar)$ and $\exp(-i\ln(q)\s D/\hbar)$, and we immediately introduce the affine coherent states
         \bn |p,q\>\equiv e^{\t ip\s\s Q/\hbar}\,e^{\t -i\ln(q)\s D/\hbar}\,|\eta\>\;, \en
         defined for all $(p,q)\in {\mathbb R}\times {\mathbb R}^+$, for some choice of the fiducial unit vector $|\eta\>$. We let the implicit domain of $q$ serve to distinguish the affine  coherent states from the canonical coherent states. We choose the fiducial unit vector as (the analog of) an extremal weight vector for the affine Lie algebra. In particular, we adopt
           \bn  [\s (Q-1)+i\s D/(\beta\s\hbar)\s]\,|\eta\>=0\;, \en
           which directly implies that $\<\eta|\s Q\s|\eta\>=1$ and $\<\eta|\s D\s|\eta\>=0$.
           If we diagonalize $Q>0$ as $x>0$, and set $D=-i\s\half\s\hbar[x\s(\d/\d x)+(\d/\d x)\s x]$, we find that
           \bn [\s \beta\s(x-1)+x\s \d/\d x+\half\s]\,\eta(x)=0\;,\en
           which leads to
           \bn \eta(x)=M\,x^{\beta-1/2}\,e^{\t-\beta\s x} \en
            with $M$ a normalization factor. Thus the $x$-representation of the affine coherent states becomes, for $x>0$,
              \bn \<x|p,q\>=M\,x^{-1/2}\,(x/q)^\beta\, e^{\t ip\s x/\hbar}\,e^{\t -\beta\s(x/q)}\;, \en
        and the affine coherent state overlap function is given by
          \bn \<p',q'|p,q\>\hskip-1.4em&&=\int_0^\infty \!\<p',q'|x\>\<x|p,q\>\s dx\no\\
          &&=\bigg\{\!\frac{(q'\s q)^{-1/2}}{[\half(q'^{-1}+q^{-1})+i\half(\beta\hbar)^{-1}(p'-p)]}\!\bigg\}^{2\s\beta}\;,\en
          using normalization to eliminate $M$. These states also admit a resolution of unity for the Hilbert space of interest given by the relation
            \bn \one=\tint |p,q\>\<p,q|\,dp\s\s dq/2\pi\hbar\s C \en
            integrated over the half plane, where $C\equiv\<\eta|\s Q^{-1}|\eta\>=1/(1-1/2\beta)$, provided that $\beta>1/2$. The Fubini-Study metric for the affine coherent states is given by
         \bn d\sigma(p,q)^2=[\beta\s\hbar]^{-1}\, q^2\s dp^2+[\beta\s\hbar]\,q^{-2}\s dq^2\;, \en
which has a constant curvature of $-2/[\beta\s\hbar]$.

            Instead of regarding $\hbar$ and $\beta$ as independent parameters above, it is sometimes very useful to
            regard $\hbar$ and ${\tilde\beta}\s\s[\equiv \beta\s\hbar]$ as independent parameters, particularly when considering limits in which $\hbar\ra0$. This alternative notation can be introduced when appropriate.

            We now turn to dynamics and the enhanced classical theory. Let $\H(P,Q)$ $=\H'(D,Q)$ denote the Hamiltonian operator, and note that the action functional for Schr\"odinger's equation is determined  by the quantum action functional
             \bn A'_Q=\tint_0^T\,\<\psi(t)|[\s i\hbar\d/\d t-\H'(D,Q)]\s|\psi(t)\>\,dt\;, \en
             which on restriction to the affine coherent states becomes
             \bn A'_{Q(R)}\hskip-1.4em&&=\tint_0^T\<p(t),q(t)|[\s i\hbar\d/\d t-\H'(D,Q)]|p(t),q(t)\>\,dt\no\\
                &&=\tint_0^T[-q(t)\s {\dot p}(t)-H(p(t),q(t))\s]\,dt\;, \en
                where $\<p,q|\s d|p,q\>=\<\eta|[\s-dp\s\s qQ-d\s \ln(q)\s D\s]|\eta\>=-q\s dp$ and
                \bn H(p,q)=\<\eta|\H(P/q+p,qQ)|\eta\>=\<\eta|\H'(D+p\s q\s Q,q\s Q)|\eta\>\;. \en
                The truly classical Hamiltonian is then given by
                  \bn H_c(p,q)=\lim_{\hbar\ra0}\,H(p,q)\;. \en

 For the case of the canonical coherent states, we argued that a classical (i.e., macroscopic) observer could only vary the position and the velocity of the system; instead, for the affine coherent states we may argue that the classical observer can only vary the (de)magnification and the velocity of the system.

            We emphasize that these results show, according to the enhanced quantization viewpoint, {\it the classical limit of an affine quantization leads to a traditional classical canonical theory
            just as much as the classical limit of a canonical quantization!}

                In the next section, Sec.~2, we consider a toy model that has singularities for all classical solutions with positive energy, but the enhanced classical theory as determined by the
                correct choice of coherent states is such that all classical solutions become nonsingular.

 \subsection{Features of the enhanced quantization formalism}
There are several important lessons to take away from the discussion up to this point. The standard formulation of canonical quantization makes a genuine conceptual leap by ``promoting'' $c$-number, phase-space, canonical (Cartesian) coordinates $p$ and $q$ to Hermitian operators $P$ and $Q$, which satisfy the canonical commutation relation $[Q,P]=i\s\hbar \s\one$. Thus, there is an implicit linkage between $q$ and $Q$ as well as between $p$ and $P$. If the canonical phase space coordinates are
changed, the corresponding operators should change accordingly; indeed, there are some who claim that unitary transformations of the operators reflect canonical coordinate transformations at the classical level. On the contrary, in the enhanced quantization procedures developed above, the classical variables $p$ and $q$ lie, effectively, in a space {\it dual} to
that of $Q$ and $P$, as seen in the arguments of the unitary transformations of the fiducial vector, i.e.,
$\exp(-iqP/\hbar)\,\exp(ip\s Q/\hbar)$, that form the canonical coherent states. Each of these factors is a one-parameter unitary group expressed in so-called canonical group coordinates. It is this dual formulation
that allows us to find a connection through their action functionals that permits both the classical and quantum stories to coexist within a single action functional distinguished only by their distinct domains.  Additionally, this approach seems to favor a specific kind of connection of the quantum Hamiltonian and the enhanced classical Hamiltonian when $\hbar>0$. This can lead to $\hbar$-modified classical equations of motion, i.e., enhanced equations of motion, that may, on the one hand, have little effect on the set of solutions to those equations, or, on the other hand, may have a very dramatic effect on the set of solutions to those equations. Examples of such dramatic changes are the subject of later sections in this article.

\subsection{Additional enhanced quantum examples}
  Enhanced quantization, wherein variables regarded as ``classical'' are regarded as dual to the ``quantum'' variables can be widely applied, even beyond traditional canonical and affine variables, and in this short subsection
  we sketch two such examples on how to formulate matters in a way such that the ``classical'' and ``quantum'' stories coexist.

  \subsubsection{A brief discussion of spin variable quantization}
 The foregoing type of connection between enhanced classical variables and associated quantum operators  is not limited to canonical and affine variables, and we take a brief detour to illustrate a similar story for spin variables. We let $S_1,\s S_2,$ and $S_3$ denote traditional spin operators with spin value $s$ as determined by the
requirement that $\Sigma_{l=1}^3\s S_l^2=\hbar^2 s(s+1)$, $s\in\{\s 1/2,1,3/2,2,\ldots\s\}$. The spin operators
with spin $s$ have an irreducible representation in a $(2\s s+1)$-dimensional Hilbert space. The operator $S_3$ has normalized eigenvectors and eigenvalues that satisfy $S_3\s |s,m\>=\hbar\s m\s |s,m\>$, where $-s\le m \le s$. We choose $|s,s\>$, an extremal weight vector for which $(S_1+i S_2)\s|s,s\>=0$,  as our
fiducial vector, and define the spin coherent states by
    \bn   |\theta,\p\>\equiv e^{\t-i\p\s S_3/\hbar}\,e^{\t -i\theta  S_2/\hbar}\,|s,s\> \en
    for all $0\le\theta\le\pi$ and $-\pi<\p\le\pi$ (as on the unit sphere).
    Observe that up to a phase, these vectors are defined by the full group of spin transformations since,
    using Euler coordinates,
     \bn |\theta,\p\>= e^{\t i\psi\s s}\,e^{\t-i\p\s S_3/\hbar}\,e^{\t -i\theta  S_2/\hbar}\,e^{\t-i\psi S_3/\hbar}|s,s\>\;. \en
These states satisfy many properties, but our interest here is to focus on the the enhanced classical/quantum connection. Let the Hamiltonian operator be $\H=\H({\bf S})$, and observe that the action functional for the Schr\"odinger equation is again given by
  \bn A_Q=\tint_0^T\<\psi(t)|\s[\s i\hbar \d/\d t-\H\s]\s|\psi(t)\>\,dt\;. \en
  Restricting the domain of allowed vectors to the set of coherent states leads to the restricted quantum action functional
   \bn  A_{Q(R)}=\tint_0^T\s\<\theta(t),\p(t)|\s[\s i\hbar \d/\d t-\H\s]\s|\theta(t),\p(t)\>\,dt\;, \en
   and it follows that this latter expression becomes
     \bn A_{Q(R)}=\tint_0^T\s[\s s\s\hbar\s \cos(\theta(t))\s\s {\dot\p}(t)-H(\theta(t),\p(t))\s]\,dt\;. \en
     The resultant expression determines the enhanced classical equations of motion; indeed, the spin action functional $A_{Q(R)}$ itself would vanish completely if we were to let $\hbar\ra0$. The spin story sketched here offers a clear example of the dual nature between the classical and quantum variables along with the coexistence of the enhanced classical and quantum formulations.

     We can also recast the spin case into canonical language, by letting $p\equiv (s\s\hbar)^{1/2}\s\cos(\theta)$ and $q\equiv (s\s\hbar)^{1/2}\s \p$, variables which are restricted such that $-(s\s\hbar)^{1/2}\le p\le (s\s\hbar)^{1/2}$ and $-\pi\s (s\s\hbar)^{1/2}< q\le \pi\s (s\s\hbar)^{1/2}$. Under this change of variables, it follows that
        \bn A_{Q(R)}=\tint_0^T\s[\s p(t)\s{\dot q}(t)-H(p(t),q(t))\s]\,dt\;. \en
     Thus, a spin-variable quantization is one way to quantize a classical canonical system with a phase space of finite area, $4\s\pi \s (s\hbar)$ [related to the Fubini-Study metric for a sphere of radius $(s\hbar)^{1/2}$], and
     resulting in a Hilbert space with $(2\s s+1)$-dimensions, $s\ge1/2$.

\subsubsection{Extension of enhanced quantization to second quantization}
Although this example is framed in the language of a canonical set of variables, it also applies
to the other previous examples, namely, to affine variables and spin variables.

Second quantization, which primarily has only historical interest at the present time, arises when the
wave function of first quantization $\<x|\psi\>$ and its adjoint $\<\psi|x\>$ are themselves turned into an annihilation  operator $A(x)$ and a creation operator $A(x)^\dag$, respectively, which satisfy $[A(x),A(x')\s]=0$ and $[A(x),A(x')^\dag]=\delta(x-x')\s\one$, and
  which act on vectors $|\Psi)$ in an appropriate Hilbert space ${\bf H}$; in particular, there is a ``no-particle'' state $|{\bf 0})$ such that $A(x)\s|{\bf 0})=0$ for all $x$ (hence, yielding a Fock representation). A second quantized action functional that captures this story is given by
  \bn A_{Q^2}= \tint_0^T(\Psi(t)|\s\{\s\tint\!\tint[ A(x)^\dag \<x|[i\hbar\s\d/\d t-\H]|x'\>\,A(x')\s]\,dx\s\s dx'\s\}\,|\Psi(t))\,dt \;.\en
  However, if we introduce standard canonical coherent states, with complex arguments $\<x|\psi\>$ and  $\<\psi|x\>$, given by
   \bn |\s|\psi\>\s) \equiv \exp\{\tint [\s A(x)^\dag\<x|\psi\>-\<\psi|x\>\,A(x)\s]\,dx\}\,|{\bf 0})\;, \en
   and use the standard property that $A(x)\s|\s|\psi\>\s)=\<x|\psi\>\s|\s|\psi\>\s)$, it follows that
   \bn A_{Q^2(R)}\hskip-1.3em&&=\tint_0^T(\<\psi(t)|\s|\s\{\s\tint\!\tint[ A(x)^\dag \<x|[i\hbar\s\d/\d t-\H]|x'\>\,A(x')\s]\,dx\s\s dx'\s\}\,|\s|\psi(t)\>)\,dt \no\\
      &&=\tint_0^T\s\<\psi(t)|[i\hbar\d/\d t-\H]|\psi(t)\>\,dt=A_Q\;. \en
   Hence, the enhanced classical limit of a second quantized system corresponds to a first quantization;
   and thus a  second quantization---or even a third or fourth, etc., quantization---of a first quantized system offers nothing new. We do not pursue this topic any further.

\section{Enhanced Quantization of a Toy Model}
     The solutions of certain classical systems often develop singularities in a finite
     time, illustrating perhaps some apparent fault with the underlying problem. However, it is quite possible that the quantum theory may not exhibit any singularities if the system is quantized in an enhanced manner. It may also be possible in those cases that the enhanced classical theory, i.e., the classical theory augmented by suitable quantum corrections in which $\hbar>0$, leads to the removal of the former singularities of the solutions. Although it would be significant if that could be the case for classical gravity which does tend to exhibit singularities, it is much easier to see if the principle of this idea even works for a simple, toy model, as we now discuss. In this section we
     rely on \cite{acsIOP,askl}.

\subsection{Toy model: the classical story}
             Let us discuss a simple example of a dynamical system where $q>0$. Consider the classical action functional given by
              \b A=\tint[\s -q\s{\dot p}-q\s p^2\s]\,dt\;,\e
              which we regard as a toy model of classical gravity for which $q(t)>0$ represents the metric quantity $\sqrt{-g}\s{ g}^{\mu\s\nu}(x)$ with its signature constraints and $p(t)$ represents minus the Christoffel symbol $\Gamma^\a_{\beta\s\gamma}(x)$.
              %see, e.g., \cit{}, which we follow closely.
              The equations of motion that follow from this toy model
              are given by
                \b {\dot q}=2\s p\s q\;, \hskip3em {\dot p}=-p^2\;, \e
                which have the general solutions
                 \b p(t)=p_0(1+p_0\s t)^{-1}\;, \hskip3em q(t)=q_0\,(1+p_0\s t)^2\;, \e
                 where $(p_0,q_0)\in({\mathbb R},{\mathbb R}^+)$ are initial values at $t=0$.
                 Clearly, {\it every} solution with positive energy, i.e., $E_0=q_0\s p_0^2>0$, exhibits
                 a singularity at $t=-1/p_0$; only if $p(t)=p_0=0$ and $q(t)=q_0>0$ are there no singularities.

                 Let us see if quantum corrections to the classical equations can possibly eliminate these singularities.

\subsection{Toy model: canonical quantization}
Let us reexamine the toy model of gravity introduced above. We start with a canonical quantization in which the Hamiltonian becomes $H=p\s q p\ra\H=P\s Q\s P$, and also introduce canonical coherent states $|p,q\>$ as defined previously with $|\eta\>=|0\>$. In that case, the enhanced  classical Hamiltonian  is given by
                  \b H(p,q)\hskip-1.3em&&=\<p,q|\s P\s Q\s P\s|p,q\>\no\\
                  &&=\<0|\s (P+p)(Q+q)(P+p)\s|0\>\no\\
                  &&=q p^2+q\s\<0|P^2|0\>\no\\
                  &&\equiv q p^2+a^2 q\;,  \e
                  where $a^2=\half\hbar$.   Thus, the enhanced equations of motion are given by
                   \b {\dot q}=2\s p\s q\;, \hskip3em {\dot p}=-(p^2+a^2)\;, \e
                   which lead to  the solutions
                  \b p(t)=a\cot(a\s (t+\a))\;,\hskip3em q(t)=(E_0/a^2)\s \sin(a\s (t+\a))^2 \;, \e
                  where  $E_0=q_0(p_0^2+a^2)>0$, and these solutions exhibit
                   singularities for {\it all} energies (since $q_0>0$ is required). Thus an enhanced canonical quantization of this toy model has not helped in removing singularities.
\subsection{Toy model: affine quantization}
                Next, let us reexamine the same toy model using an enhanced affine quantization in which the Hamiltonian becomes $H=p\s q p=d\s q^{-1} d=H'\ra \H'=D\s Q^{-1} D$, and also introduce  affine coherent states instead of canonical coherent states. For affine coherent states, recall that $\<\eta|D|\eta\>$=0 and $\<\eta|Q|\eta\>=1$, and moreover, $\<\eta| Q^p|\eta\>=1+O(\hbar)$, for all $p\ge-1$, adopting $\hbar$ and ${\tilde\beta}\s\s[=\beta\s\hbar]$ as independent parameters; this property  implies that $Q$ (and $q$) are dimensionless, while the dimensions of $P$ (and $p$) are those of $\hbar$. In this case, the enhanced classical Hamiltonian is given by
                    \b H(p,q)\hskip-1.3em&&=\<p,q|\s DQ^{-1}D\s|p,q\>\no\\
                    &&=\<\eta|\s (D+p\s q\s Q)(q\s\s Q)^{-1}(D+p\s q\s Q)\s|\eta\>\no\\
                      &&=q p^2+ \<\eta|\s D\s Q^{-1}D\s|\eta\>\,q^{-1}\no\\
                      &&\equiv q p^2+\hbar^2\s C\s q^{-1}\;, \e
                     for some dimensionless constant $C>0$. It is already clear from the form of the enhanced Hamiltonian that {\it all} solutions to the enhanced equations of motion with finite energy are nonsingular because
                      \b E_0=q(t)\s p(t)^2+\hbar^2\s C\s q(t)^{-1}\ge \hbar^2\s C\s q(t)^{-1}\;. \e
                      In this case the extended equations of motion become
                        \b  {\dot q}=2\s p\s q\;, \hskip3em {\dot p}=-p^2+\hbar^2 C\s q^{-2}\;, \e
                        and the new solutions are given by
                    \b p(t)=\frac{(t+\a)}{{(t+\a)^2+\hbar^2\s C/4E_0^2}} \;,\hskip2em
                    q(t)=4E_0[ (t+\a)^2+\hbar^2\s C/4E_0^2]\;,\e
                    where $E_0=q_0p_0^2+\hbar^2\s C/q_0>0$.

                  It is noteworthy that adopting the enhanced quantization point of view enables us to consider an affine quantization as well as a canonical quantization. Based on the primacy of self-adjoint kinematical variables, the affine approach is favored, and using the associated affine
                   coherent states leads to an enhanced classical Hamiltonian that eliminates singularities in solutions of the enhanced classical equations of motion, singularities that were present in the original classical theory.

                   In a recent paper \cite{zon}, Fanuel and Zonetti  carry out a similar study for certain simplified models of an initial cosmological singularity, and, at least for the few models studied, they
                  conclude that the enhanced quantization viewpoint and an affine quantization that was required did indeed remove singularities in the usual classical solutions.

\section{Enhanced Quantization of \\Rotationally Symmetric Models}
           The nonlinear classical model under consideration in this section is formulated for both a finite ($N<\infty$) and an infinite ($N=\infty$) number of degrees of freedom. When $N<\infty$, the quantization procedure is straightforward and acceptable. However, when $N=\infty$, a natural limit ($N\ra\infty$) of the quantization procedures of the nonlinear
           classical Hamiltonian leads to a {\it free} quantum theory. To resolve this problem requires us to exploit a feature of enhanced quantization other than affine quantization. In this section we closely follow \cite{kiev,bookBCQ,rotsym}.

         Consider the classical Hamiltonian for an $N$-degree of freedom problem given by
          \b H(\op,\oq)=\half [\op^2 +m^2_0\oq^2]+ \lambda_0(\oq^2)^2\;, \e
          where $\op\equiv\{p_n\}_{n=1}^N$ and $\op^2\equiv\op\cdot\op=\Sigma_{n=1}^N \s p_n^2$ as well as $\oq\equiv\{q_n\}_{n=1}^N$ and  $\oq^2\equiv\oq\cdot\oq=\Sigma_{n=1}^N \s q_n^2$, where $N\le\infty$. When $N<\infty$,
          there is no restriction on the values of $\{p_n\}$ and $\{q_n\}$; when $N=\infty$, the values of $\{p_n\}$ and $\{q_n\}$ must satisfy  $\op^2+\oq^2<\infty$. These models
          have a rotational symmetry that is very helpful in our analysis, a feature that also accounts for the name of the models. Given $\op$ and $\oq$, there are three rotationally invariant variables: $X\equiv \op^2$, $Y\equiv\op\cdot\oq$, and
          $Z\equiv\oq^2$. The energy is $E=\half X+\half m^2_0 Z + \l_0 Z^2$, $\l_0\ge0$, and the magnitude of the angular momentum is determined by $\overrightarrow{L}^2=(\op\times\oq)^2=X\s Z-Y^2\ge0$.
          The solution to the equations of motion is equivalent to a {\it one}-dimensional problem ($N=1$) if
          $\overrightarrow{L}=0$, while the solution is equivalent to a {\it two}-dimensional problem ($N=2$) if $\overrightarrow{L}\not=0$. It is important to keep this equivalence in mind in order to see if that property is also shared by the quantum theory.

\subsection{Free model: excluding the nonlinear interaction}
          In a straightforward canonical quantization, the variables $\op\ra\oP$ and $\oq\ra\oQ$ such that $[Q_j,P_k]=i
          \hbar\delta_{j, k}\one$. For  the free theory, where $\l_0=0$ and for simplicity letting $m_0\ra m$, the operator $\H_f\equiv\half[\oP^2+m^2\oQ^2]$ would be satisfactory for $N<\infty$, but when $N=\infty$, we need to subtract the zero-point energy, which may be done by normal ordering, leading to $\H_0\equiv\half:[\oP^2+m^2\oQ^2]:\s$.  To make the transition $N\ra\infty$ smoother, we will adopt normal
ordering for $N<\infty$ as well.

          To understand the effects of normal ordering (all creation operators to the left of all annihilation operators), it is helpful to introduce the annihilation operators
          $a_n\equiv (\sqrt{m} Q_n$ $+i\s P_n/\sqrt{m})/\sqrt{2\hbar}$ and the creation operators $a_n^\dag\equiv(\sqrt{m} Q_n-i\s P_n/\sqrt{m})/\sqrt{2\hbar}$, which obey $[a_n,a_m]=0$ and $[a_n,a^\dag_m]=\delta_{n,m}\one$. In that
          case $\H_0=\half:[\oP^2+m^2\oQ^2]:\,=(m\s\hbar)\Sigma_{n=1}^N\s a^\dag_n a_n$ which is clearly a well-defined operator for $N\le\infty$. Reinforcement of its operator character is given by
           \bn \H_0^2=[\s (m\hbar)\Sigma_{n=1}^N\s a^\dag_n a_n\s]^2=\,:[\s (m\hbar)\Sigma_{n=1}^N\s a^\dag_n a_n\s]^2:+(m\hbar)^2\Sigma_{n=1}^N a^\dag_na_n\;, \en
           which is also well defined.

          Canonical coherent states are also used to study this example, and they are defined by $|\op,\oq\>=\exp(-i\oq\cdot\oP/\hbar)\s \exp(i\op\cdot\oQ/\hbar)\s|0\>\equiv U[\op,\oq]\s|0\>$.
               Here, the normalized fiducial vector $|0\>$  satisfies $(m\oQ+i\oP)\s|0\>=0$ or equivalently
               $(m\oQ-i\oP)\s\cdot\s(m\oQ+i\oP)\s|0\>=\;:\oP^2+m^2\oQ^2:\s|0\>=0$. Thus,
               it follows that
               \bn H_0(\op,\oq)\equiv \<\op,\oq|\H_0|\op,\oq\>=\half[\op^2+m^2\oq^2]\;,\en
               and, as an example of the good operator properties of $\H_0$, we note that
                \bn \<\op,\oq|\H_0^2|\op,\oq\>=\{\s\half[\op^2+m^2\oq^2\s]\}^2+\half(m\hbar)[\op^2+m^2\oq^2]\;.\en

\subsection{True model: including the nonlinear interaction}
               We now study the model when $\l_0>0$. The nonlinear interaction term is given by $:(\oQ^2)^2:$
               which has the diagonal coherent state matrix elements  $\<\op,\oq|:(\oQ^2)^2:|\op,\oq\>=(\oq^2)^2$. However, this acceptable relation conceals the fact that, as an operator, $:(\oQ^2)^2:={\cal O}(N)$, as demonstrated by the property that $[\s:(\oQ^2)^2:\s]^2$ has terms proportional to $\Sigma_{l,m,n,p}\, a^2_la^2_ma^{\dag\,2}_na^{\dag\,2}_p\propto N^2+l.o.t.$ (here $l.o.t.=$ lower order terms). For very large $N$, this fact is commonly dealt with by rescaling the coupling constant, i.e., $\l_0\ra\l_0/N$, leading to the revised Hamiltonian operator
                 \bn \H'=\half:[\oP^2+m^2_0\oQ^2]:+(\l_0/N):(\oQ^2)^2:\;.\en
                 For $1\ll N<\infty$, it is common to study similar problems by a perturbation series in $1/N$ \cite{oneoverN}. As a method to evaluate what happens for very large (but finite) $N$, the $1/N$-expansion may be appropriate, but we are really interested in the ultimate case where $N=\infty$. Thanks to the new form of the coupling constant, it follows, when $N\ra\infty$, that the nonlinear Hamiltonian becomes  a free theory again, with a possibly changed mass ${\widetilde m}$, given by $\H'_0=\half:[\oP^2+{\widetilde m}^2\oQ^2]:\,$. Thus, when $N=\infty$, and the nonlinear classical model is well defined for $\l_0>0$, it turns out
                 that the conventional quantization procedure leads to a {\it free} quantum theory, a quantum theory that has as its own classical limit a {\it free classical model} that is definitely  not the {\it interacting classical model} we started with.

                 Surely, this should not be the end of the story!

     \subsection{Invoking the weak correspondence principle}
                 How can we overcome this unsatisfactory situation? As part of the enhanced quantization program, we start by appealing to the Weak Correspondence Principle \cite{WCP} which asserts that
                   \bn H(\op,\oq)\equiv \<\op,\oq|\s\H\s|\op,\oq\>\;. \en
                   Assuming---as is traditional in conventional quantization procedures---that the representation of $\oP$ and $\oQ$ is irreducible leads to the situation recounted above. But, according to the weak correspondence principle, the assumption of irreducible basic kinematical operators is {\it NOT} required! Instead, as a substitute for irreducibility, it is logical to assume that the coherent states $|\op,\oq\>$ completely span the Hilbert space of interest, even if the operators $\oP$ and $\oQ$ are now reducible. This new proposal opens the door to an alternative narrative!

                   The new proposal holds for both $N=\infty$ and $N<\infty$, and, for simplicity, let us assume that $N<\infty$ unless otherwise stated.
               For the free theory, where $\l_0=0$, the operator $\H_0=\half:[\oP^2+m^2\oQ^2]:$ is well defined and, as developed above, it corresponds to the classical free Hamiltonian $H_0(\op,\oq)=\half[\op^2+m^2\oq^2]\s$; this
               assignment is valid for $N=\infty$ as well. The fiducial vector $|0\>$ is also the ground state for $\H_0$,
               namely, $\H_0\s|0\>=0$, and in the Schr\"odinger representation, $\<\ox|0\>=N'\s\exp\{-(m/2\hbar)\s\ox^2\}$. In addition, the operator
                 $ \H_1\equiv \H_0+\gamma :\H_0^2: $
                 is also well defined for $N\le\infty$, and with $|\op,\oq\>=U[\op,\oq]\s|0\>$, it follows that
                    \b H_1(\op,\oq)\equiv \<\op,\oq|\H_1|\op,\oq\>=\half[\op^2+m^2\oq^2]+\gamma\s\{\half[\op^2+m^2\oq^2]\}^2\,.
                    \no\\  \e
                    This expression contains $(\oq^2)^2$, but it also contains unwanted terms; nevertheless, that fact offers a clue regarding how to proceed.

                    We now make explicit the appropriate reducible operator representation based on the careful study in \cite{rotsym}.
                    Besides the set of operators $\oP$ and $\oQ$, we introduce a second, independent set of similar operators, namely, $\oR$ and $\oS$ with the commutation relation $[S_j,R_k]=i\hbar\delta_{j,k}\one$; as usual, independence means that any operator from $(\oP,\oQ)$ commutes with any operator from $(\oR,\oS)$. The new Hamiltonian operator is taken to be
                    \b &&\H_2\equiv\half:[\oP^2+m^2(\oQ+\zeta\s\oS)^2]:+\half:[\oR^2+m^2(\oS+\zeta\s\oQ)^2]:\no\\
                    &&\hskip8.4em+\beta:[\oR^2+m^2(\oS+\zeta\s\oQ)^2]^2:\;, \e
                    where $\zeta$ is a new parameter with $0\le\zeta<1$. Each of the basic components, i.e.,
                    $:[\oP^2+m^2(\oQ+\zeta\s\oS)^2]:$ and $:[\oR^2+m^2(\oS+\zeta\s\oQ)^2]:$, is like a
                    free field Hamiltonian, much as was $:[\oP^2+m^2\oQ^2]:\s$. Just as the vector $|0\>$ is the ground state of the latter operator, the vector $|0,0;\zeta\>$ is the ground state for both of the former operators, and in the Schr\"odinger representation, $\<\ox,\oy|0,0;\zeta\>=N''\s\exp\{-(m/2\hbar)[\ox^2+\oy^2+2\zeta \ox\cdot\oy]\}.$ The coherent states of interest are given by
                    $|\op,\oq\>\equiv U[\op,\oq]\s|0,0;\zeta\>$, and in the Schr\"odinger representation we have
                    \b&&\hskip-1.8em\<\ox,\oy|\op,\oq\>\\
                      &&\hskip-1.4em=N''\exp\{i\op\cdot(\ox-\oq)/\hbar-(m/2\hbar)[(\ox-\oq)^2+\oy^2
                    +2\zeta(\ox-\oq)\cdot\oy]\}\,.\no\e
                     For $\zeta>0$, the functions $\<\ox,\oy|\op,\oq\>$ span
                     a space equivalent to the Schr\"odinger  space $L^2({\mathbb R}^{2N})$; however, if we set $\oq=0$, then the functions $\<\ox,\oy|\op,0\>$ do {\it not} span the same space, and instead only span a space equivalent to the Schr\"odinger  space $L^2({\mathbb R}^N)$. In this sense, with the right choice of the fiducial vector, coherent states made from a reducible representation of the basic operators span a larger space that is {\it not} spanned by coherent states made from an irreducible representation of the basic operators. To complete the story we observe that
                     \bn H_2(\op,\oq)\hskip-1.3em&&\equiv\<\op,\oq|\s\H_2\s|\op,\oq\>\no\\
                       &&=\half[\op^2+m^2(1+\zeta^2)\s\oq^2]+\beta\s m^4\zeta^4\s(\oq^2)^2\no\\
                        &&\equiv\half[\op^2+m^2_0\s\oq^2]+\l_0\s(\oq^2)^2\;. \en
                        Note that this desirable classical/quantum connection requires that $\zeta>0$ in order to include the nonlinear interaction, a fact that also ensures a reducible representation of the basic operators.

                        With this last expression, we have demonstrated a formulation of the quantum theory that completely
                        resolves the Rotationally Symmetric models, and, significantly,  some features of the spectrum of this model have been discussed in \cite{TAYLOR}. When the number of degrees of freedom $N=\infty$, the free theory is the only case covered by irreducible representations
                        of the basic operators. To treat the interacting cases when $N=\infty$, it is essential to use reducible representations, and, as a consequence, the correct solution for interacting models is compatible with the principles of enhanced quantization.
                        Moreover, the new quantum formulation has an identical analytic formulation whether $N<\infty$ or $N=\infty$, which is similar to the properties of the classical model.
 \section{Enhanced Quantization of \\Ultralocal Scalar Models}
                 The models discussed in this section are nonrenormalizable as well as trivial
                 when analyzed by conventional quantization procedures. In turn, these unsatisfactory results are overturned when enhanced quantization procedures are introduced. The final
                 results turn out to be completely satisfactory. In this section we are guided by
                 \cite{acsIOP,bookBCQ,ultrascalar}.
 \subsection{Canonical quantization of ultralocal models}
                      Consider the classical phase-space action functional
                      \b A_0=\tint\{\s\pi(t,\bx)\s{\dot \p}(t,\bx) -\half\s[\s\pi(t,\bx)^2+m_0^2\s\p(t,\bx)^2\s]\s\}\,dt\s d^s\!x\;,\e
                      which describes the free ultralocal scalar field model where $x\in{\mathbb R}^s$, $s\ge1$. This model differs from a relativistic  free theory---which is discussed in Sec.~6---by the absence of the term $[\overrightarrow{\nabla}\p(t,\bx)]^2$, and therefore, for the ultralocal models, the
                      temporal behavior of the field at one spatial point $\bx$ is independent of the temporal behavior of the field at any spatial
                      point $\bx'\not=\bx$.  Traditional canonical quantization of the classical Hamiltonian,
                      i.e., $\pi(\bx)\ra\hp(\bx)$ and $\p(\bx)\ra\hph(\bx)$ where  $H(\pi,\p)\ra\H=H(\hp,\hph)$, leads to an infinite ground state energy, so it is traditional to proceed differently. Formally speaking, the classical Hamiltonian density
                      is first reexpressed as
                       \b \half[\pi(\bx)^2+m_0^2\s\p(\bx)^2]=\half[m_0\p(\bx)-i\s\pi(\bx)][m_0\p(\bx)+i\s\pi(\bx)]\;,\e
                       and only then is it quantized directly. This procedure leads to the quantum Hamiltonian
                       \b \H_0\hskip-1.3em&&=\half\int[m_0\hph(\bx)-i\s\hp(\bx)][m_0\hph(\bx)+i\s\hp(\bx)]\,d^s\!x\no\\
                       &&=\half\int[\hp(\bx)^2+m_0^2\s\hph(\bx)^2-\hbar\s m_0\s\delta(0)]\,d^s\!x\no\\
                       &&\equiv\half\int\s:[\hp(\bx)^2+m_0^2\s\hph(\bx)^2]:\,d^s\!x\;, \e
                       a result which leads to conventional normal ordering symbolized, as we have already done, by  $:\,:\s$.

                       If we regularize this expression by a finite, $s$-dimensional, hypercubic, spatial lattice,  and adopt a Schr\"odinger representation, then the Hamiltonian operator
                       becomes
                       \b \H_0=\half\S_k[-\hbar^2 \s a^{-2s}\d^2/\d\p_k^2+m^2_0\s\p_k^2-\hbar\s m_0\s a^{-s}]\,a^s\;. \e
                       The primed sum $\Sigma'_k$ denotes a sum over the spatial lattice, where $k=(k_1,\ldots,k_s)\in{\mathbb Z}^s$, labels
                       sites on the spatial lattice.  In this equation
                       $a>0$ is the lattice spacing, $a^s$ is an elementary spatial cell volume, and on the lattice we
                      have used  $\hp(\bx)\ra-i\hbar\s a^{-s}\d/\d\p_k$ and $\hph(\bx)\ra \p_k$. The ground state of this Hamiltonian is just the
                       product of a familiar Gaussian ground state,
                         \b \psi_k(\p_k)=(m_0\s a^s/\pi\hbar)^{1/4}\, e^{\t - m_0\p_k^2\s a^s/2\hbar}\;,\e
                       for a large number of independent, one-dimensional harmonic oscillators, and thus
                      the characteristic functional (i.e., the Fourier transform) of the
                      ground-state distribution is given by
                        \b C_0(f)\hskip-1.3em&&=\lim_{a\ra0}N_0\int e^{\t (i/\hbar)\S_k\s f_k\s\p_k\, a^s-(m_0/\hbar)\S_k\p_k^2\, a^s}\,\Pi'_k d\p_k\no\\
                            &&= {\cal N}_0\int e^{\t (i/\hbar)\tint f(\bx)\s\p(\bx)\,d^s\!x-(m_0/\hbar)\tint \p(\bx)^2\,d^s\!x}\,\Pi'_\bx d\p(\bx)\no\\
                            &&=e^{\t -(1/4\s m_0\hbar )\tint f(\bx)^2\,d^s\!x}\;. \e
                      The primed product $\Pi'_k$ runs over all sites on the spatial lattice.
                     In the last two lines of this relation the continuum limit has been taken (with a formal
                     version in the second line)
                     in which $a$ goes to zero, $L$, the number of sites on each edge, goes to infinity, but $a\s L$ remains finite (at least initially).

                     Now, as just one example, suppose we introduce a quartic nonlinear interaction leading
                     to the classical phase-space action functional,
                       \b A=\tint\{\pi(t,\bx)\s{\dot\p(t,\bx)}-\half\s[\s\pi(t,\bx)^2+m_0^2\s\p(t,\bx)^2\s]\s
                       -g_0\s\p(t,\bx)^4\s\}\,dt\s\s d^s\!x \;. \e
                       Again, the temporal development of the field at one spatial point is independent
                       of the temporal development at any other spatial point.
                     Using the same lattice regularization as for the free theory and introducing
                     normal ordering for the interaction, the characteristic functional of the ground-state distribution is necessarily of the form
                       \b C(f)\hskip-1.3em&&=\lim_{a\ra0}N\int e^{\t (i/\hbar)\S_k\s f_k\s\p_k\,a^s-\S_k {\widetilde Y}(\p_k,g_0,\hbar,a)\,a^s}\,\Pi'_k d\p_k\no\\
                       &&= {\cal N}\int e^{\t (i/\hbar)\tint f(\bx)\s\p(\bx)\,d^s\!x-\tint Y(\p(\bx),g_0,\hbar)\,d^s\!x}\,
                       \Pi'_\bx d\p(\bx)\no\\
                       &&=e^{\t -(1/4\s{\sf m}\hbar)\tint f(\bx)^2\, d^s\!x}\;.\e
                       Here ${\widetilde Y}(\p_k,g_0,\hbar,a)$ and $Y(\p(\bx),g_0,\hbar)$  denote some non-quadratic functions that arise in the solution of the nonlinear Hamiltonian ground-state differential equation on the lattice and likewise in the formal
                       continuum limit, respectively. Importantly,  {\it the last line is a consequence of the Central Limit Theorem} \cite{CLT}, yielding a {\it free theory} with a positive mass {\sf m}, a factor that absorbs
                       all trace of the quartic interaction. In short, a conventional canonical quantization of this nonlinear quantum field theory---which is a nonrenormalizable quantum field theory since all closed loops in a perturbation analysis diverge when integrated over spatial momentum---has rigorously led to a ``trivial'' (free) theory, even though the original classical theory was nontrivial.

                       The question naturally arises: Can we change quantization procedures to yield a nontrivial result?

\subsection{Affine quantization of ultralocal models}
                       We start by modifying the classical Hamiltonian before quantization. This modification is completely different from the one used for the conventional quantization discussed above.
                       For both $g_0=0$ and $g_0>0$, let us consider
                       \b \hskip-2em H(\pi,\p)\hskip-1.3em&&=\tint\{\s\half[\pi(\bx)^2+m_0^2\s\p(\bx)^2]+g_0\s\p(\bx)^4\}\,d\bx\no\\
                       &&=\tint\{\s\half[\pi(\bx)\p(\bx)\p(\bx)^{-2}\p(\bx)\pi(\bx)+m_0^2\s\p(\bx)^2]
                       +g_0\s\p(\bx)^4\}\,d\bx\;.\label{ee4}\e
                       If the classical Poisson bracket for the field $\p(x)$ and the momentum $\pi(x')$
                       is given by $\{\p(x),\pi(x')\}=\delta(x-x')$, then the Poisson bracket between the
                       field $\p(x)$ and the dilation field $\rho(x')\equiv \pi(x')\s\p(x')$ is given by
                       $\{\p(x),\rho(x')\}=\delta(x-x')\s\p(x)$.
                       On quantization, we will treat the classical product $\pi(\bx)\p(\bx)$ as the dilation field $\rho(\bx)$ and we will invoke affine quantum commutation relations for which
                       $\p(\bx)\ra\hph(\bx)$ and $\rho(\bx)\ra \hr(\bx)$ such that $[\hph(\bx),\hr(x')] =i\hbar\s\delta(\bx -x')\s\hph(\bx)$. Note well: if we choose affine commutation relations for which $\hph(\bx)$ and $\hr(\bx)$, when smeared, are self-adjoint operators, then it follows that the canonical momentum operator $\hp(\bx)$, when smeared, is only a {\it form}  and {\it not} an operator due to the local operator product involved; recall that a form requires restrictions on kets {\it and} bras. In short, when quantizing fields, one can choose either affine field variables or canonical field variables, since both systems generally cannot exist simultaneously
                       because of the local operator product involved.

                       As compared to Sec.~1, where affine variables were introduced, it is noteworthy that
                       the affine field variables $\hph(x)$ and $\hr(x)$ we presently deal with comprise a reducible representation of operators since the spectrum of $\hph(x)$, when smeared with a positive test function, runs over the whole real line. However, the
                       choice of the fiducial vector we make restricts the space spanned by the affine coherent states to a subspace equivalent to an irreducible representation.

                          To see what affine quantization leads to, let us work formally and focus on twice
                          the kinetic energy density. Thus, classically,
                          \b \pi(\bx)^2\hskip-1.3em&&=\pi(\bx)\p(\bx)\p(\bx)^{-2}\p(\bx)\pi(\bx)\no\\
                      &&=\rho(\bx)\p(\bx)^{-2}\,\rho(\bx)\;, \e
                      which on quantization becomes
                      \b \hr(\bx)\s\hph(\bx)^{-2}\hr(\bx)\hskip-1.3em&&=\quarter[\hp(\bx)\hph(\bx)+\hph(\bx)\hp(\bx)]
                      \hph(\bx)^{-2}[\hp(\bx)\hph(\bx)+\hph(\bx)\hp(\bx)]\no\\
                      &&=\quarter[2\hp(\bx)\hph(\bx)+i\hbar\delta(0)]\hph(\bx)^{-2}[2\hph(\bx)\hp(\bx)-i\hbar\delta(0)]\no\\
                      &&=\hp(\bx)^2+i\s\half\hbar\s\delta(0)[\hph(\bx)^{-1}\hp(\bx)-\hp(\bx)\hph(\bx)^{-1}]\no\\
                      &&\hskip6em +
                         \quarter \hbar^2\delta(0)^2\hph(\bx)^{-2}\no\\
                         &&=\hp(\bx)^2+\threequarters\hbar^2\delta(0)^2\hph(\bx)^{-2}\;. \e
                         The factor $\threequarters$ will have an essential role to play.

                         Guided by this calculation we introduce the same lattice regularization to again quantize the classical free theory (i.e., $g_0=0$), but this time focussing on an affine quantization, namely
                         \b \H'_0=\half\S_k[-\hbar^2 a^{-2s}\d^2/\d\p_k^2+m_0^2\p_k^2+F\s\hbar^2 a^{-2s}  \p_k^{-2}-E'_0\s]\,a^s\;, \e
                         where $F\equiv(\half-b\s a^s)(\threebytwo-b\s a^s)$; here $b>0$ is a constant factor with dimensions (Length)$^{-s}$, and $E'_0$ is explained below. Note that $F$ is a regularized form of $\threequarters$, and it becomes that number in the continuum limit. Again the ground state
                         of this Hamiltonian is a product over one-dimensional ground states for each
                         independent degree of freedom, and for a small lattice spacing, each ground state wave function,  has the form
                           \b \psi_k(\p_k)=(b\s a^s)^{1/2}\,e^{\t - m_0\phi_k^2\,a^s/2\hbar}\,|\p_k|^{-(1-2ba^s)/2}\;,\label{ww4}\e
                           with a ground-state energy of $E'_0=\half\s\hbar\s m_0 \s ba^s$; observe that the
                           ground-state energy on the lattice is finite  and that it{\it vanishes} in the continuum limit! This new form of ground-state wave function corresponds to what we have called a ``pseudofree''
                           model \cite{bookBCQ}, and we see that it also arises upon quantization of the free classical model by exploiting an unconventional factor-ordering ambiguity and insisting on securing affine kinematical variables, $\hph(\bx)$ and $\hr(\bx)$.

                           In effect, as the continuum limit is approached, the new factor in (\ref{ww4}) serves to ``mash the measure'' (see Sec.~6 for a fuller explanation of this phrase) in the sense that under a change of $m_0$ the {\it free} ground-state distributions are {\it mutually singular} (with disjoint support) while the {\it pseudofree} ground-state distributions are {\it equivalent} (with equal support). This very fact can be seen as the reason for the divergence-free character
                       of the affine quantization for these models.

                           The characteristic functional for the pseudofree ground-state distribution is given by
                           \b C_{pf}(f)\hskip-1.3em&&=\lim_{a\ra0}\int \Pi'_k\{ (ba^s)\,e^{\t i f_k\p_k\s a^s/\hbar-m_0\p_k^2\s a^s/\hbar}\,|\p_k|^{-(1-2ba^s)}\s\}\,\Pi'_k d\p_k\no\\
               &&=\lim_{a\ra0}\Pi'_k\s\{\s 1- (b\s a^s)\tint[1-\cos(f_k\p\s a^s/\hbar)]\,e^{\t -m_0\p^2\s a^s/\hbar}\,d\p/|\p|^{(1-2ba^s)}\s\}\no\\
                           &&= \exp\{-b\tint d\bx\tint[1-\cos(f(\bx)\s\l/\hbar)\s]\,e^{\t-b\s m\s\l^2/\hbar}\,d\l/|\l|\}\;,\label{w13}\e
                           where $m_0\equiv ba^s\s m$ and $\l\equiv\p\s\s a^s$; note the multiplicative renormalization involved in $m_0=b\s a^s\s m$. The final distribution described by (\ref{w13}) is a generalized Poisson distribution \cite{LUK}. It is instructive to become convinced that only the  number  $\threequarters$ could have led to  this desirable result.

                           Furthermore, it is also noteworthy that the pseudofree Hamiltonian that
                           annihilates the pseudofree ground state wave function has a discrete spectrum with {\it uniform spacing} as well as a {\it vanishing} zero-point energy \cite{bookBCQ}.

                           Finally, we consider the affine quantization of an ultralocal model with a quartic interaction as described by the classical action in (\ref{ee4}). In that case the lattice form of the quantum Hamiltonian acquires the additional term
                           $g_0\S_k\p_k^4\s a^s$ [with no normal ordering needed, but rather $g_0=(b\s a^s)^3\s g$, with $g\ge0$] along with a suitable change of $E'_0$. Now the lattice regularized
                           ground-state wave function at each site is, for small values of $a$,  of the form
                            \b \psi_k=(ba^s)^{1/2} \,e^{\t-\half\s {\widetilde y}(\p_k,\hbar,a)\s a^s}\,|\p_k|^{-(1-2ba^s)/2}\e
                            for some appropriate nonquadratic function ${\widetilde y}(\p_k,\hbar,a)$. In turn, the characteristic functional of the interacting ground-state distribution becomes, in the continuum limit,
                            \b C(f)=\exp\{-b\tint d\bx\tint[1-\cos(f(\bx)\s\l/\hbar)\s]\,e^{\t -b^{-1}\s y(b\l,\hbar)}\,
                            d\l/|\l|\}\;, \e
                       where ${\widetilde y}(\p,\hbar,a)$ is related to $b^{-1}y(b\l,\hbar)$ by $\l=\p\s\s a^s$ and multiplicative renormalization of suitable coefficients. General arguments from the theory of infinite divisibility \cite{LUK} show that only the free (Gaussian) or the nonfree (generalized Poisson)
                       distributions, with characteristic functions for ground state distributions
                       illustrated above, are allowed by
                       ultralocal symmetry.

                       Note that following the route of affine quantization for the ultralocal model has resulted in overcoming the triviality of quantization that conventional canonical procedures invariably lead to for this model. We also observe that suitable affine coherent states have been
                       central in the quantum/classical connection for ultralocal models; see \cite{bookBCQ}.
                       Other approaches exist that lead to the same conclusions we have obtained in this section, but in keeping with the general theme of this article, we have focussed on an affine variable approach as part of the program of enhanced quantization.

                       In Sec.~6 we continue to discuss scalar fields and we analyze {\it covariant} scalar fields using
                       tools closely related to those used in the present section to resolve the ultralocal models. But first we tackle quantum gravity.

\section{Enhanced Quantization of Gravity}
Unifying classical gravity and quantum theory is a very difficult problem, and current approaches include
superstring theory and loop quantum gravity. Although these approaches have achieved much, there is still plenty of work to be done. In what follows we offer a sketch of the Affine Quantum Gravity program pioneered by the author. In a certain sense, quantum gravity is an ideal theory for an affine quantization approach!
In this section we are guided by \cite{SLAV} and references therein.

\subsection{Virtues of the classical affine variables}
    In canonical phase-space coordinates, the classical formulation of Einstein's theory of gravity in a  $3+1$ spacetime may be described by the  action functional
   \b &&A=\int [\s\pi^{a b}(t,x)\s{\dot g}_{ab}(t,x)-N^a(t,x)\s H_a(\pi,g)(t,x)\no\\
   &&\hskip4em -N(t,x)\s H(\pi,g)(t,x)\,]\,dt\s\s d^3\!x\;. \e
   Here $\pi^{ab}\s(=\pi^{b\s a})$ is the momentum, $g_{ab}\s(=g_{b\s a})$ is the metric, with $a,b=1,2,3$, and
   $N^a(t,x)$ and $N(t,x)$ are Lagrange multipliers to enforce, respectively, the diffeomorphism constraints  $H_a(\pi,g)(t,x)=0$ and
   the Hamiltonian  constraint $H(\pi,g)(t,x)=0$. The range of the momentum $\pi^{ab}(t,x)$ is all of ${\mathbb R}$ for each component,
   while the range of the spatial metric $g_{ab}(t,x)$ is such that the $3\times3$ matrix $\{g_{ab}(t,x)\}$
   is positive definite for all $(t,x)\in {\mathbb R}^4$; stated otherwise, if $u^a$ is an arbitrary, nonvanishing vector, then $u^a\s g_{ab}(t,x)\s u^b>0$. {\it Note well: The insistence on positive definiteness of the spatial matrix $\{g_{ab}(t,x)\}$ for all $(t,x)$ is a fundamental principle of
   the affine formulation of both classical and quantum gravity}.
   This criterion implies the existence of an inverse metric tensor $g^{bc}(t,x)$, also positive definite, for which $g_{ab}(t,x)\s g^{bc}(t,x)=\de^c_a$. The classical equations of motion follow from stationary variation of the action and they can be expressed in terms of Poisson brackets as
    \b &&{\dot g}_{ab}(t,x)=\{g_{ab}(t,x), H_T\}\;, \hskip3em {\dot \pi}^{ab}(t,x)=\{\pi^{ab}(t,x), H_T\}\;,\no\\
       &&\hskip3.36em H_a(\pi,g)(t,x)=0\;, \hskip3em H(\pi,g)(t,x)=0\;. \e
       Here $H_T=\tint [\s N^a(t,x)\s H_a(\pi,g)(t,x)+N(t,x)\s H(\pi,g)(t,x)\s]\,d^3\!x$, and the
       fundamental Poisson bracket among the basic canonical variables at equal times (and now dropping the time dependence) is given by
       \b \{\s g_{ab}(x), \pi^{cd}(y)\}=
       \half[\s\de^c_a\s\de^d_b+\de^d_a\s\de^c_b\s]\,\de(x,y)\;. \e

       As dictated by the constraints, some aspects of the 12 phase-space variables at each point $x$ are unphysical.
       The Poisson brackets of the constraints among themselves all vanish on the constraint hypersuface, i.e,  that region of phase space defined as ${\cal C}\equiv\{(\pi,g): \s H_a(\pi,g)(x)=0,\s H(\pi,g)(x)=0\s\}$, and they are then classified as first-class constraints. It follows
       that once the
       phase-space variables $\pi^{ab}(x)$ and $g_{ab}(x)$ initially lie on the constraint hypersurface, say at $t=0$,
        then, as time passes, the phase-space variables remain on the constraint hypersurface for any
       choice of the Lagrange multipliers. Stated otherwise, the equations of motion do not determine the
       Lagrange multipliers, which is physically natural since the Lagrange multipliers set the temporal coordinate distance to the next spatial surface as well as the location on that surface of the spatial coordinates themselves, all variables that may be freely specified. However, the constraints do not form a (formal) Lie algebra since the structure ``constants'' are
       actually functions of the phase-space coordinates, a situation referred to as open first-class constraints. The constraints play a crucial role for gravity as they clearly determine all the relevant physics.

       Unlike the case of scalar field quantization, the dilation variable suitable for classical relativity is a familiar quantity, namely, $\pi^a_c(x)=\pi^{ab}(x)\s g_{bc}(x)$; it is also very easy to retrieve the original canonical momentum since $\pi^{ac}(x)=\pi^a_b(x)\s g^{b c}(x)$. The set of Poisson brackets satisfied by the classical affine variables $\pi^a_b(x)$ and  $g_{ab}(x)$
       at equal times is given by
          \b &&\{\pi^a_b(x),\pi^c_d(y)\}=\half[\de^c_b\,\pi^a_d(x)-\de^a_d\,\pi^c_b(x)]\,\de(x,y)\;,\no\\
  && \hskip-.08cm\{g_{ab}(x),\pi^c_d(y)\}=\half[\de^c_a\,g_{db}(x)+\de^c_b\,g_{ad}(x)]\,\de(x,y)\;,\\
  && \hskip-.18cm\{g_{ab}(x),g_{cd}(y)\}=0\;,\no  \label{e1}\e
  which follows directly from the fundamental Poisson bracket between $g_{ab}(x)$ and $\pi^{cd}(y)$
  given above. As was previously the case, the Poisson brackets for the affine variables form a (formal) Lie algebra.
  It is clear that one can reformulate the classical theory
  of gravity entirely in terms of these affine variables. Note well that $\pi^a_b$ has {\it nine}
  independent components while $g_{ab}$ has {\it six} independent components. Thus, there is no way in which the dilation variables $\pi^a_b$ can be considered as one-to-one ``partners'' of the metric variables $g_{ab}$ as canonical variables would be.
  To see the ``dilation'' aspects offered by $\pi^a_b$, consider the macroscopic canonical transformation induced by the arbitrary, nine-component tensor $\g^a_b(y)$ coupled together with $\pi^b_a(y)$ and with
  $\pi(\g)\equiv\tint\s\g^a_b(y)\s\pi^b_a(y)\,d^3\!y$ as the generator, namely
      \b &&\hskip-2em e^{\t\s\{\,\cdot\, ,\s\pi(\g)\s\}}\;g_{cd}(x)\no\\
       &&\equiv g_{cd}(x)+\{g_{cd}(x),\pi(\g)\}+\half \{\{g_{cd}(x),\pi(\g)\},\pi(\g)\}+ \cdots\no\\
       &&\equiv M^a_c(x)\, g_{ab}(x)\s M^b_d(x)\;, \e
   %    where we have set
    %    $\pi(\g)\equiv\tint\s\g^a_b(y)\s\pi^b_a(y)\,d^3\!y$
        where the components of the matrix $M(x)$ are
         \b M^a_b(x)\equiv\{e^{\t \half\g(x)\s}\}^a_b\e
           while $\g(x)$ is the $3\times3$ matrix $\{\g^a_b(x)\}$. Clearly, if $\{g_{ab}\}$ is a positive definite matrix
         before the transformation, it remains a positive definite matrix after the transformation. Even
         though the word ``dilation'' does not capture all the properties of the matrix $M^a_b(x)$, we shall still refer to the affine variable $\pi^a_b(x)$ as the dilation variable. [{\bf Remark:} Occasionally,  we have used the word ``momentric''
         in referring to $\pi^a_b(x)$, a name that is derived from the two words {\it momen}tum and me{\it tric}.]

         A macroscopic canonical transformation with ${\widetilde\pi}(\chi)\equiv \tint
         \chi_{ab}(y)\s\pi^{ab}(y)\,d^3\!y$ as generator would lead to $g_{cd}(x)+\chi_{cd}(x)$, a result that
         may transform a positive definite metric into a non-positive definite metric, thus potentially violating the fundamental principle of metric positivity.

\subsection{Quantum aspects of the affine variables}
         Our discussion of an affine quantization of gravity focusses on the use of affine variables for which $\pi^a_b(x)\ra\hp^a_b(x)$ and $g_{ab}(x)\ra \hg_{ab}(x)$, where these local operators are required to fulfill the affine commutation relations given by \b &&[\hp^a_b(x),\hp^c_d(y)]=i\half\s\hbar\s[\de^c_b\,\hp^a_d(x)-\de^a_d\,\hp^c_b(x)]\,\de(x,y)\;,\no\\
  &&\hskip-.1em[\hg_{ab}(x),\hp^c_d(y)]=i\half\s\hbar\s[\de^c_a\,\hg_{db}(x)+\de^c_b\,\hg_{ad}(x)]\,\de(x,y)\;,\\
  &&\hskip-.3em[\hg_{ab}(x),\hg_{cd}(y)]=0\;,\no  \label{e1}\e
  which are seen to be a direct transcription of the Poisson brackets for the classical affine variables. Observe that the affine commutation relations form a (formal) Lie algebra, and have the appearance of current commutation relations, which leads to the fact that there are
  generally very different operator representations than those that arise in the case of canonical variables which satisfy canonical commutation relations.

  The choice of a representation for the affine operators is largely determined by choosing a suitable fiducial vector for a set of coherent states. In our case, we introduce
    \b |\pi,\g\>\equiv e^{\t (i/\hbar)\tint \pi^{ab}(y)\s\hg_{ab}(y)\,d^3\!y}\,e^{\t -(i/\hbar)\tint \g^a_b(y)\s \hp^b_a(y)\,d^3\!y}\,|\eta\>\;, \e
    and look for arguments to choose a suitable fiducial vector.
    To proceed further, we need to introduce some physics. Dirac's quantization procedure for systems with constraints requires quantization first and reduction second  \cite{DIR}.
    In brief, we are asked to quantize initially as if there were no constraints at all; after quantization, we are then required to restrict the original Hilbert space $\frak H$ to the physical Hilbert space ${\frak H}_{phys}$ by some means.
    Therefore, we initially focus on a quantization without regard for the constraints, which means taking a neutral position regarding how field values at one spatial point are related to field values at any other spatial point. The neutral and natural way to do so is to assume that the initial operator representation is ultralocal in character. This means, for example, that the proper  coherent state overlap function
    must have the form
      \b \<\pi'',\g''|\pi',\g'\>\equiv e^{\t-\tint L[\pi''(x),\g''(x);\pi'(x),\g'(x)]\,d^3\!x} \;.\e
      If we imagine regularizing this expression by a spatial lattice, then we can find the form of $L$ by studying what happens at a single lattice site. This study has been carried out in \cite{aqg1} by
      making a choice of the fiducial vector that---like the elementary one-dimensional affine example studied in Sec.~1---is, effectively, annihilated by a complex linear combination of the affine variables (i.e, the fiducial vector is basically an extremal weight vector). The result of that study leads to
      \b  \<\pi'',\g''|\pi',\g'\>\hskip-1.3em&&= \exp\bigg(\!-\!2{\int} b(x)\,d^3\!x\, \no\\
  &&\hskip-3em\times\ln\bigg\{  \frac{
\det\{\half[g''^{kl}(x) +g'^{kl}(x)]+i\half [\hbar\s b(x)]^{-1}[\pi''^{kl}(x)-
\pi'^{kl}(x)]\}} {(\det[g''^{kl}(x)])^{1/2}\,(\det[g'^{kl}(x)])^{1/2}}
\bigg\}\bigg)\no\\
  && \equiv\<\pi'',g''|\pi',g'\> \;.\label{et3}\e

Observe that the matrices $\gamma''$ and $\gamma'$ do {\it not} explicitly appear in this expression; symmetry properties within the chosen
 $|\eta\>$ are such that the matrix $\{\gamma^a_b\}$ has  been replaced by the
 positive definite matrix $\{g_{ab}\}$, where
  \b  g_{ab}(x)\equiv M_a^c(x)\,\<\eta|\hg_{cd}(x)|\eta\>\,M_b^d(x)\;, \e
      and, as before, the matrix elements $M_a^c(x)=\{e^{\t\half\gamma(x)}\}_a^c$. Since symmetry of the fiducial vector has reduced  dependence of the coherent states from the nine components of $\g^a_b$ to the six components of $g_{ab}$, we are free to relabel the coherent states as $|\pi,g\>\equiv |\pi,\g\>$ to reflect that symmetry, and we have already labeled the coherent state overlap $\<\pi',g''|\pi',g'\>$ in anticipation of that fact. In this expression,
      the function $b(x)>0$---also determined by $|\eta\>$---is a scalar density with dimensions (Length)$^{-3}$, and thus we observe that the given coherent state overlap function is {\it invariant} under arbitrary spatial coordinate transformations of $\pi$, $g$, and $b$.
      Moreover, the physical meaning of the affine coherent state labels $\pi^{ab}(x)$ and $g_{ab}(x)$ follows from the fact that
         \b \<\pi,g|\s\hg_{ab}(x)\s|\pi,g\>\hskip-1.3em&&=g_{ab}(x)\;, \no\\
            \<\pi,g|\s\hp^a_c(x)\s|\pi,g\>\hskip-1.3em&&=\pi^{ab}(x)\s g_{bc}(x)\equiv \pi^a_c(x)\;. \e
      The coherent state overlap is a continuous function of positive type, and therefore it can serve as a reproducing kernel for a natural functional representation of the original Hilbert space $\frak H$ by continuous functionals \cite{aron}. Indeed,  in this section, a reproducing kernel Hilbert space is our preferred choice for defining functional element representatives and the inner product of such elements.

      Note that the metric ${\tilde g}_{ab}(x)\equiv\<\eta|{\hat g}_{ab}(x)|\eta\>$ serves only to characterize the topology of the space-like surface $\Sigma$ over which the integral in (\ref{et3}) is carried out.

      Since the coherent state overlap function is, apart from normalization factors,
       analytic in its arguments, i.e., $g^{ab}(x)\pm i\pi^{ab}(x)/b(x)$, it follows
      that the affine coherent state overlap function for gravitational affine variables may
      be given a functional integral representation of the form
      \b  && \<\pi'',g''|\pi',g'\>  \no\\
 &&\hskip1cm=\lim_{\nu\ra\infty}{\o{\cal N}}_\nu\s\int
e^{\t-(i/\hbar)\tint[g_{ab}\s\s{\dot\pi}^{ab}]\,dt\s d^3\!x}\no\\
  &&\hskip-1cm\times\exp\{-(1/2\nu)\tint[\s\s[\hbar\s b(x)]^{-1}g_{ab}\s g_{cd}\s
{\dot\pi}^{bc}\s{\dot\pi}^{da}+[\hbar\s b(x)]\s g^{ab}\s g^{cd}\s{\dot g}_{bc}\s{\dot g}_{da}\s]\,
\,dt\s d^3\!x\}\no\\
  &&\hskip2cm\times[\Pi_{t,x}\,\Pi_{a\le b}\,d\pi^{ab}(t,x)\,
dg_{ab}(t,x)]\;. \label{fe8} \e
This functional integral representation is {\it not} postulated; instead, it is {\it derived} from the analytic nature of the coherent state overlap function.
    Observe the implicit metric on phase space functionals that is part
   of the $\nu$-dependent, continuous-time regularization factor in (\ref{fe8}); this metric acts to force all histories of $g_{ab}(t,x)$ in the functional integral to respect metric positivity. Although this functional integral has all the appearances of a phase-space functional integral, it is important to keep in mind that this formulation arose from an {\it affine} quantization procedure and the functional integral yields the overlap function for {\it affine} coherent states.

\subsection{Enforcing the gravitational constraints}
    So far we have not included the effect of the constraints in accordance with
    the Dirac approach to the quantization of systems with constraints \cite{DIR}, where the rule is to quantize first and to reduce second. The reason behind this rule is that by reducing first, there is the strong chance that the remaining classical phase space may not admit Cartesian coordinates leading to an
    uncertain quantization procedure. By choosing to quantize first, one can usually
    arrange a physically secure quantization procedure. After the initial step of quantization, Dirac advocates generating the physical Hilbert space from the original Hilbert space by asking that the constraint operators all give zero when acting on those vectors that make up the subspace of
    the original Hilbert space that forms the physical Hilbert space. This procedure works for constraints
    that are elements of a compact Lie algebra. However, this procedure encounters problems with constraints for which zero lies in the continuous spectrum as well as for all constraint systems that are second class (defined by Poisson brackets or commutators of constraints that do not vanish on the constraint hypersurface or on the physical Hilbert space, respectively).

    A more recent variation of Dirac's procedure  \cite{proj} instead builds a projection operator $\E$ that projects onto a small spectral subspace, between zero and $\delta(\hbar)^2$ (where  $\de(\hbar)$ is a regularization parameter), of the sum of the squares of the constraint operators, a method that treats both first- and second-class constraints on an equal footing. The regularized  physical Hilbert space (regularized by $\de(\hbar)$) is then
     defined by ${\frak H}_{phys}=\E\s\s{\frak H}$. To enforce the constraints properly, it is necessary to reduce the regularization parameter $\de(\hbar)$ to its smallest optimal size, which can either be zero (for first-class constraints) or nonzero (for partially second-class constraints). If the spectrum of the sum of the squares of the constraint operators near zero lies in the continuum, then it is necessary to take a suitably rescaled  form limit of the projection operator $\E$ as $\de\ra0$. This limit results in a reproducing kernel that leads to a Hilbert space on which the constraints are properly enforced.

     Advantages of the projection operator method include:
     no gauge fixing, a path integral formulation with no auxiliary variables, no need for Dirac brackets or other means to eliminate second-class constraints, as well as several others. Since quantization of constraints is not one of our primary topics, we have only offered a very brief summary of the projection operator method of dealing with systems with constraints; the interested reader can learn much more about this method from \cite{proj,proj2,sch,aqg2}.

     Even though we have offered few details, we have sketched enough information so that the reader
     will be able to sense the meaning of an introduction of the constraints in a regularized fashion.
     In this discussion, we seek the affine coherent state matrix elements of the regularized projection operator $\E$ by means of a specially weighted functional integral involving the Lagrange multiplier functions,
     $N^a(t,x)$ and $N(t,x)$, and the constraint symbols, ${\tilde H}_a={\tilde H}_a(\pi,g)(t,x)$ and ${\tilde H}={\tilde H}(\pi,g)(t,x)$ (symbols because $\hbar>0$), along with a suitable measure, $R(N^a,N)$, which is deliberately designed to enforce the {\it quantum} constraints and {\it not the  classical} constraints. The resultant functional integral is formally a modest generalization of the functional integral for the affine coherent state overlap, and is given by
      \b  && \<\pi'',g''|\s\E\s|\pi',g'\>  \no\\
 &&\hskip1cm=\lim_{\nu\ra\infty}{\o{\cal N}}_\nu\s\int
e^{\t-(i/\hbar)\tint[g_{ab}\s\s{\dot\pi}^{ab}+N^a\s{\tilde H}_a+N\s{\tilde H}]\,\,dt\,d^3\!x}\no\\
  &&\hskip-1cm\times\exp\{-(1/2\nu)\tint[\s[\hbar\s b(x)]^{-1}g_{ab}\s g_{cd}\s
{\dot\pi}^{bc}\s{\dot\pi}^{da}+[\hbar\s b(x)]\s g^{ab}\s g^{cd}\s{\dot g}_{bc}\s{\dot g}_{da}]\,dt\,
d^3\!x\}\no\\
  &&\hskip2cm\times[\Pi_{t,x}\,\Pi_{a\le b}\,d\pi^{ab}(t,x)\,
dg_{ab}(t,x)]\,\D R(N^a,N)\;. \label{e8} \e
This expression can serve as a reproducing kernel for a functional representation of the regularized physical Hilbert space ${\frak H}_{phys}$ by continuous functionals.
For further details regarding the program of affine quantum gravity, the interested reader is referred to
\cite{SLAV} and references therein. More recent  results can be found in \cite{newAQG}.

\section{Enhanced Quantization of \\Covariant Scalar Models}
\subsection{Nature of the problem \& conventional wisdom}
As a representative example of the models considered in this section, we focus on a scalar field $\p(t,x)$, $x\in{\mathbb R}^s$, $s\ge1$, with the
classical action functional
  \bn A=\tint\{\half[{\dot\p}(t,x)^2-(\overrightarrow{\nabla}\p)(t,x)^2-m_0^2\s\p(t,x)^2]-g_0\s\p(t,x)^4\s\}\,
  dt\,d^s\!x\;, \en
  as well as the imaginary-time classical action functional, now with $x\in{\mathbb R}^n$, $n=s+1$,
\bn I=\tint\{\half[(\nabla\p)(x)^2+m_0^2\s\p(x)^2]+g_0\s\p(x)^4\s\}\,d^n\!x\;. \label{e1}\en
 Such models have been extensively studied by a variety of conventional techniques including perturbation theory, renormalization group studies, Monte Carlo computations, and continuum limits of lattice regularized, imaginary time, functional integral formulations, just to mention a few approaches. The conventional wisdom has settled on the view that for the superrenormalizable models ($n=2,3)$ all traditional methods
point to a well-behaved quantization that reduces to the original, nonlinear classical model when $\hbar\ra0$. On the other hand, for
the strictly renormalizable model ($n=4$) as well as all of the nonrenormalizable models ($n\ge5$), the consensus is that the quantum theory becomes a free (or generalized free) model which as $\hbar\ra0$ necessarily converges to a free (or generalized free) classical model that does
{\it not} agree with the original, nonlinear classical model. We have seen a similar behavior for a technically simpler problem in Sec.~4 when we studied the ultralocal models. In particular,
conventional analysis of the ultralocal models led to a nonrenormalizable perturbation analysis as well as a lattice formulation that passed to a free quantum theory in the continuum limit, and thus also passed to a free model in the classical limit. However, in the case of the ultralocal models, by appealing to an affine quantization procedure rather than a conventional canonical quantization, we were able to determine
a different, divergence free and acceptable outcome for those models that were nontrivial and passed to the original, nonlinear model in the classical limit. Our goal in this section is to  outline a similar study for covariant scalar models.

However, before we begin, we need to acknowledge several strong prevailing opinions that attempt to explain and even justify the
less-than-desirable results of conventional quantization such as triviality for $n\ge4$. One argument asserts that the poor results
for such models are the fault of the models themselves and not of the quantization procedure. Some of these models
are nonrenormalizable (requiring an infinite number of distinct counterterms), and as one story goes,
nature hasn't chosen those models because they involve infinitely many unknown parameters. Another claim is that
the contribution of nonrenormalizable models is small and some as-yet-unknown theory (often envisioned to be superstring theory) will resolve any problems encountered at high energy by nonrenormalizable theories. In the same spirit, such troublesome models are deemed to qualify only as ``effective theories'' valid at low energies and for a few terms in a perturbation analysis. Monte Carlo studies have shown quite convincingly that the $\p^4_4$ model becomes free in the continuum limit \cite{wei}, and there are rigorous proofs for $\p^4_n$, $n\ge5$, that the $\p^4_n$ models become free in the continuum limit \cite{aiz,fro}. {\it These results are not wrong}. Rather they are based {\it solely} on the
consequences of a quantization based on a standard canonical procedure, and they have not dealt with the counterterm that we propose. Indeed, as we have argued throughout this
article, there is another quantization procedure for classical models that we have called affine quantization. Focusing on affine quantization gave us acceptable results for the ultralocal models. Let us see if similar techniques have something positive to say about covariant $\p^4_n$ models!

In the effort to offer something new and better, we are guided by \cite{IOP,CCURE}.

\subsection{Highlights of an affine quantization}       % Affine quantization of $\p^4_n$ models
Let us present our basic results right at the beginning. Much as we were led to a lattice Hamiltonian for the ultralocal models with an unconventional counterterm, now, for the covariant models, we
are led to a lattice Hamiltonian on a finite, periodic, hypercubic lattice with lattice spacing $a>0$, and where $k^*$ runs over the $s$ spatially nearest neighbor sites to $k$ in the positive direction,  given by
 \bn \label{WeH} \H\hskip-1.1em&&=-\half\hbar^2\s a^{-2s}{\t\sum}'_k\,\frac{\d^2}{\d\p_k^2}\,a^s
     +\half \hbar^2\s a^{-2s}{\t\sum}'_k\frac{1}{\Psi_0(\p)}\,\frac{\d^2\s\Psi_0(\p)}{\d\p_k^2}\,a^s\no\\
    &&=-\half\hbar^2\s a^{-2s}{\t\sum}'_k\,\frac{\d^2}{\d\p_k^2}\,a^s+\half{\t\sum}'_{k,k^*}(\p_{k^*}-\p_k)^2\s a^{s-2} \\
    &&\hskip2em+\half\s m_0^2{\t\sum}'_k\p_k^2\,a^s+
    g_0{\t\sum}'_k\p_k^4\,a^s
    +\half\hbar^2{\t\sum}'_k{\cal F}_k(\p)\,a^s-E_0\;, \no\en
    in which the all-important counterterm ${\cal F}_k(\p)$ is given by
    \bn \label{WeF} {\cal F}_k(\p)
                \hskip-1.4em&&=\quarter\s(1-2ba^s)^2\s
          a^{-2s}\s\bigg({\ts\sum'_{\s t}}\s\frac{\t
  J_{t,\s k}\s \p_k}{\t[\Sigma'_m\s
  J_{t,\s m}\s\p_m^2]}\bigg)^2\\
  &&\hskip-1em-\half\s(1-2ba^s)
  \s a^{-2s}\s{\ts\sum'_{\s t}}\s\frac{J_{t,\s k}}{[\Sigma'_m\s
  J_{t,\s m}\s\p^2_m]}
  +(1-2ba^s)
  \s a^{-2s}\s{\ts\sum'_{\s t}}\s\frac{J_{t,\s k}^2\s\p_k^2}{[\Sigma'_m\s
  J_{t,\s m}\s\p^2_m]^2}\;. \no \en
  The latter two equations implicitly determine the ground-state wave function $\Psi_0(\p)$, which has the general form
  \bn \Psi_0(\p)=M' e^{\t-U(\p,\hbar,a)/2}\,\Pi'_k[\Sigma'_l\s J_{k,l}\s\p^2_l]^{-(1-2ba^s)/4}.\label{Ww3}\en
   Here $J_{k,l}\equiv 1/(2 s+1)$ for $l=k$ and for $l$ equal to each of the $2\s s$ nearest neighbors to $k$ in a spatial sense; otherwise $J_{k,l}\equiv 0$. Thus $\Sigma'_l\s J_{k,l}\s\p^2_l$ is an
  average of $(2s+1)$ field-squared values at the site $k$ and nearby.
   With $\Psi_0(\p)$  as the ground state wave function, it follows that  $\exp[-U(\p,\hbar,a)/2]$ is determined by the {\it large field behavior} of the Hamiltonian potential while the remaining factor $\Pi'_k[\Sigma'_l\s J_{k,l}\s\p^2_l]^{-(1-2ba^s)/4}$ is determined by the {\it small field behavior} of the Hamiltonian potential. There is no normal ordering in $\H$; instead, all the local products involved are given by operator product expansions realized by a multiplicative renormalization of the several parameters.

  In comparing these formulas to those for the ultralocal models, we see, besides the introduction of the spatial derivative term, an analogous counterterm (\ref{WeF}) that, loosely speaking, is also ``$1/\p^2$'' in character, the only difference being that the factor $|\p_k^2|^{-(1-2ba^s)/4}$ in the ultralocal ground state, which is appropriate for the symmetry of the ultralocal model, is changed to $[\Sigma'_l\s J_{k,l}\s\p^2_l]^{-(1-2ba^s)/4}$ in the covariant model ground state reflecting the fact that
  in the covariant case fields on nearby lattice sites naturally tend to locally equilibrate (thanks to the gradient term) and this averaging serves to eliminate any nascent, nonintegrable singularity in the ground state distribution. The small field factor leads to
  the same level of ``measure mashing'' for the covariant models as held for the ultralocal models (more on this topic below). The new
  form of the lattice action follows directly from the lattice Hamiltonian operator, and thus the Euclidean lattice action (where $\Sigma_k$ is a sum over all spacetime lattice sites) is given by
  \bn  I=\half \Sigma_{k,k^*}(\p_{k^*}-\p_k)^2\,a^{n-2}+\half m_0^2\Sigma_k\p_k^2\,a^n
     +g_0\Sigma_k\p_k^4
     \,a^n+\half\Sigma_k \hbar^2{\cal F}_k(\p)\,a^n\;,\no\\
      \label{eLL}\en
     where in this expression $k^*$ refers to all $n=s+1$ nearest neighbors of the site
     $k$ in the positive direction, and the unprimed sum runs over all spacetime lattice sites.
   Thanks to measure mashing, the imaginary-time, lattice functional integral for the covariant models leads to a divergence-free power series expansion of the interaction term about the pseudofree model very much like with the ultralocal models. Moreover, a preliminary Monte Carlo study supports a positive, nonvanishing renormalized coupling constant for the $\p^4_4$ theory based on the lattice action (\ref{eLL})  \cite{stank}. However, since that particular study ended before definitive results were obtained, we certainly encourage other groups to perform additional Monte Carlo calculations for these models as well.
\subsection{An affine quantization of covariant scalar models}
 To begin at the beginning, we note that the quantization of the $\p^4_n$ model  may be addressed by the formal functional integral
  \bn S(h)\equiv{\cal M}\int e^{\t (1/\hbar)\tint h(x)\s\p(x)\,d^n\!x-(1/\hbar)[I(\p)+ F(\p,\hbar)/2]}
        \,\Pi_x \s d\p(x)\;,\no\\   \label{e2}\en
  where ${\cal M}$ is chosen so that $S(0)=1$, $h(x)$ is a smooth source function, $I(\p)$ denotes the Euclidean action, and $F(\p,\hbar)$ represents an unspecified counterterm to control divergences, which should formally vanish as $\hbar\ra0$ so that the proper classical limit formally emerges. Rather than adopting a standard version of the counterterm, we keep an open mind as we seek a counterterm that eliminates all divergences.
   Although the formal
  functional integral (\ref{e2}) is essentially undefined, it can be given meaning by first introducing a lattice regularization in which the spacetime continuum is replaced with a periodic, hypercubic lattice with lattice spacing $a>0$ and with  $L<\infty$ sites along each axis. The sites themselves are labeled by multi-integers $k=(k_0,k_1,\ldots,k_s)\in{\mathbb Z}^n$, $n=s+1$, and $h_k$ and $\p_k$ denote field values at the $k$th lattice site; in particular, $k_0$ is designated as the Euclidean time variable. This regularization results in the $L^n$-dimensional integral
  \bn &&S_{latt}(h)\equiv M\int e^{ \t(1/\hbar)Z^{-{1/2}}\Sigma_k h_k\s\p_k\,a^n}\no\\
  &&\hskip7em \times\,e^{\t-(1/2\hbar)\s[\Sigma_{k,\s k^*}(\p_{k^*}-\p_k)^2\,a^{n-2}+\s m_0^2\s\Sigma_k\p_k^2\,a^n]}\no\\
  &&\hskip7em \times\,e^{\t-(1/\hbar)\s g_0\Sigma_k\s\p_k^4\,a^n-(1/2\s\hbar)\s\Sigma_k F_k(\p,\hbar)\,a^n}\;\Pi_k\s d\p_k\;.\no\\
  \label{e4} \en
  Here we have introduced the field-strength renormalization constant $Z$. The factors $Z$, $m_0^2$, and $g_0$ are
  treated as bare parameters implicitly dependent on the lattice spacing $a$, ${k^*}$ denotes one of the $n$ nearest neighbors in the positive direction from the site $k$, and $M$ is chosen so that $S_{latt}(0)=1$.
  The counterterm $F_k(\p,\hbar)$ also implicitly depends on $a$, and
  the notation $F_k(\p,\hbar)$ means that the formal, locally defined  counterterm $F(\p,\hbar)$ may, when lattice regularized,  depend on finitely-many field values located within a small, finite region of the lattice around the site $k$.

     Since the lattice regulation has led to finitely many integrations in (\ref{e4}), it is instructive to focus on the emergence of divergences as the continuum limit is taken, which we define as $a\ra0$, $L\ra\infty$, with  $a\s L$ fixed and finite. Divergences already arise  as $L\ra\infty$ without the need for $a\s L\ra\infty$ as well; for a discussion of the limit $a\s L\ra\infty$, see \cite{IOP}.

     We now return to the lattice-regularized functional integral (\ref{e4}). In order for
     this mathematical expression to be physically relevant following a Wick rotation to real time, we impose the requirement of {\it reflection positivity}
     \cite{reflec}, which is assured if the counterterm satisfies
       \bn \Sigma_k\s F_k(\p,\hbar)\equiv\Sigma_{k_0}\s \{\Sigma'_k { F}_k(\p,\hbar)\} \en
       where the expression $\{\Sigma'_k {F}_k(\p,\hbar)\}$ involves fields all of which have the {\it same} temporal value $k_0$, but may involve several other fields at nearby sites to $k$ in spatial directions only.

       Let us next consider (\ref{e4}) limited to a
       test function supported on a {\it single} spatial slice, e.g., $h_k\equiv a^{-1}\s\delta_{k_0,0}\,f_k$, for which (\ref{e4})
       becomes \cite{IOP} equivalent to
       \bn S'_{latt}(f)=\int e^{\t Z^{-1/2}\Sigma'_k f_k\p_k\s a^s/\hbar}\,\Psi_0(\p)^2\,\Pi'_kd\p_k\;, \label{e10}\en
       where $\Psi_0(\p)$ denotes the unique, normalized, ground state wave function for the Hamiltonian operator $\H$ for this problem, expressed in the Schr\"odinger representation, with
       the property that $\H\Psi_0(\p)=0$.

       We next derive the desired modification of the conventional ground-state wave function by a different argument than was used for the ultralocal models. Expressed in hyperspherical coordinates for the $N'\equiv L^s$ sites on a spatial slice---for which $\p_k\equiv \k\s\eta_k$, $\k^2\equiv \Sigma'_k\p_k^2$, $1\equiv\Sigma'_k\eta_k^2$, $0\le\k<\infty$, and $-1\le\eta_k\le1$---(\ref{e10})
       becomes
         \bn S'_{latt}(f)\hskip-1.3em&&=\int e^{\t\k Z^{-1/2}\Sigma'_kf_k\s\eta_k\,a^s/\hbar}\,\Psi_0(\k\eta)^2\,\k^{N'-1}\s d\k
         \,2\s\delta(1-\Sigma'_k\eta_k^2)\,\Pi'_k\s d\eta_k\;.\no\\
         \en
         In the continuum limit, as $L\ra\infty$ and therefore $N'=L^s\ra\infty$, divergences will generally
         arise because, in that limit, if parameters like $m_0$ or $g_0$, are changed, the measures
         become {\it mutually singular} (with disjoint support) due to the overwhelming influence of $N'$ in the measure factor
         $\k^{N'-1}$.
         However,  we can avoid that conclusion provided the ground-state distribution $\Psi_0(\p)^2$ contains a factor that serves to ``mash the measure''.
         Specifically, we want the ground-state wave function to have the form, with $R>0$ fixed and finite,
         given by
           \bn \hskip-1em\Psi_0(\p)\hskip-1.3em&&= ~^{``} M'\,e^{\t-U(\p,\hbar,a)/2}\,\k^{-(N'-R)/2}\s^{\,"}\no\\
           &&=M' e^{\t-U(\k\eta,\hbar,a)/2}\,\k^{-(N'-R)/2}    %\no\\
           %&&\hskip4em\times
           \,\Pi'_k[\Sigma'_l\s J_{k,l}\s\eta^2_l]^{-(1-R/N')/4}\no\\
               &&=M' e^{\t-U(\p,\hbar,a)/2}\,\Pi'_k[\Sigma'_l\s J_{k,l}\s\p^2_l]^{-(1-R/N')/4}.\en
               The first line (in quotes) indicates the qualitative $\k$-behavior that will effectively mash the measure, while the second and third lines illustrate a specific functional dependence on field variables that leads to the
               desired factor. As noted earlier, we choose the constant coefficients $J_{k,l}\equiv1/(2s+1)$ for $l=k$ and for $l$ equal to each of the $2s$ spatially nearest neighbors to the site $k$; otherwise, $J_{k,l}\equiv0$.
               As part of the ground state distribution, the factor including $\Sigma'_lJ_{k,l}\s\p_l^2$ is dominant for small-field values, and its form is no less fundamental than the rest of the ground-state distribution which is determined by the gradient, mass, and interaction terms that fix the  large-field behavior. The factor $R/N'$ appears in the local expression of the small-field factor, and on physical grounds that quotient should not depend on the number of lattice sites in a spatial slice nor on the specific parameters mentioned above that define the model. Therefore, we can assume that $R\propto N'$, and so we set
                              \bn   R\equiv 2\s b\s a^s\s N'\;, \en
               where $b>0$ is a fixed factor with dimensions (Length)$^{-s}$ to make $R$ dimensionless; this
               is effectively the same choice for $R$ that was made for the ultralocal models.

               Even though the ground-state distribution diverges when certain of the $\eta_k$-factors  are simultaneously zero, these are {\it all integrable singularities} since whenever any subset of the $\{\eta_k\}$ variables are near zero, there are always fewer zero factors arising from the singularities thanks to the local averaging procedure; {\it this very fact has helped motivate the averaging procedure}. Note that only affine coherent states---and not canonical coherent states---built on the ground state wave function as the fiducial vector can properly be extended to the continuum limit \cite{IOP}.

               To obtain the required functional form of the ground-state wave function for small field values, we choose our counterterm to build that feature into the Hamiltonian. In particular, the counterterm is a specific potential term of the form
               \bn\half \Sigma'_k\s{ F}_k(\p,\hbar)\,a^s\equiv \half\hbar^2\Sigma'_k{\cal F}_k(\p)\,a^s\en
               where, with $T(\p)\equiv \Pi'_r[\Sigma'_l J_{r,l}\s\p_l^2]^{-(1-2ba^s)/4}$,
                \bn {\cal F}_k(\p)\hskip-1.1em&& \equiv\frac{ a^{-2s}}{ T(\p)}\frac{\d^2 T(\p)}{\d\p_k^2}\;,\en
               with the result given by (\ref{WeF}).
  Although the counterterm ${\cal F}_k(\p)$ does not depend only on $\p_k$, it nevertheless becomes a local potential
  in the formal continuum limit. %REFLECTION POSITIVITY

  More generally, the full Hamiltonian operator including the desired counterterm is chosen as
    \bn \H\hskip-1.3em&&
    =-\half\hbar^2\s a^{-2s}{\t\sum}'_k\,\frac{\d^2}{\d\p_k^2}\,a^s+\half{\t\sum}'_{k,k^*}(\p_{k^*}-\p_k)^2\s a^{s-2} \no\\
    &&\hskip2em+\half\s m_0^2{\t\sum}'_k\p_k^2\,a^s+
    g_0{\t\sum}'_k\p_k^4\,a^s
         +\half\hbar^2{\t\sum}'_k{\cal F}_k(\p)\,a^s-E_0\;. \label{eH}\en
    As previously noted, this latter equation implicitly determines the ground-state wave function $\Psi_0(\p)$ because $\H\s\Psi_0(\p)=0$.

    It is important to observe that no normal ordering applies to the terms in the Hamiltonian. Instead,
    local field operator products are determined by an operator product expansion \cite{IOP}. In addition, note that the ``counterterm'' $\half\hbar^2\Sigma'_k{\cal F}_k(\p)$ does {\it not} depend on any parameters of the model and
    specifically not on $g_0$. This is because the counterterm is really a counterterm
    for the {\it kinetic energy}. This fact follows because not only is $\H\s\Psi_0(\p)=0$, but then $\H^q\s\Psi_0(\p)=0$ for all integer $q\ge2$. While $[\Sigma'_k\d^2/\d\p_k^2]\s\Psi_0(\p)$ may result in  a square-integrable function, the expression  $[\Sigma'_k\d^{2}/\d\p_k^{2}]^q\s\Psi_0(\p)$ will surely not be square integrable for suitably large $q$. To ensure that $\Psi_0(\p)$
    is in the domain of $\H^q$, for all $q$, the derivative term and the counterterm must be considered together to satisfy domain requirements, hence
    our claim that the counterterm should be considered as a `renormalization' of the kinetic energy.

    Since the counterterm does not depend on the coupling constant, it follows that the counterterm remains even when
    $g_0\ra0$, which means that the interacting quantum field theory does {\it not} pass to the usual free quantum field theory as $g_0\ra0$, but instead it passes to what we have called a {\it pseudofree quantum field theory}, just as was the case for the ultralocal models. As a relevant example, consider the classical (Euclidean)
        action functional (\ref{e1}). Regarding the separate components of that expression,
        and assuming both $m_0>0$ and $g_0>0$, a multiplicative inequality \cite{russ,bookBCQ} states that
          \bn \{g_0\tint \p(x)^4\,d^n\!x\}^{1/2}\le{\tilde C}\tint[(\nabla\p)(x)^2+m_0^2\p(x)^2]\,d^n\!x\label{e333}\en
          where ${\tilde C}= (4/3)\s [\s g_0^{1/2}\s m_0^{(n-4)/2}\s]$ when $n\le4$ (the renormalizable cases), and ${\tilde C}=\infty$ when  $n\ge5$ (the nonrenormalizable cases), which in the latter case means there are
          fields for which the integral on the left side of the inequality diverges
          while the integral on the right side is finite.
          In other words,  there are different free and pseudofree classical theories when $n\ge5$. Thus, for $n\ge5$, it is reasonable to assume that the pseudofree quantum field theory is also different from the free quantum field theory.
          Moreover, the quantum models developed in this section with the unconventional counterterm provide
   viable candidates for those quantum theories normally classified as nonrenormalizable, and they do so
   in such a manner that in a perturbation analysis, expanded now about the pseudofree model, divergences do not arise because all the underlying measures are equivalent and no longer mutually singular! (A discussion of the divergence-free properties from a perturbation point of view appears in \cite{IOP}, an analysis that also determines the dependence
   of $Z$, $m_0^2$, and $g_0$ on the parameters $a$ and $N'$.)

   Since the unconventional counterterm conveys good properties to the nonrenormalizable models, it is natural to extend such good behavior to the traditionally renormalizable models $(n\le4)$ by using the unconventional counterterm for them as well.
   {\it Thus, we are led to adopt the lattice regularized Hamiltonian $\H$ (\ref{eH}), including the counterterm, for all spacetime dimensions $n\ge2$.} (Observe that a similar analysis shows the same counterterm applies to {\it all} $\p^p_n$ models, where $p\in(4,6,8,\dots)$, and $n\ge2$!)

   Additionally, the lattice Hamiltonian (\ref{eH}) also determines the lattice Euclidean action (\ref{eLL}) including the unconventional counterterm.  Although the lattice form of the counterterm involves averages over field-squared values
     in nearby spatial regions of the central site, it follows that the continuum limit of the counterterm
     is local in nature, as noted previously. We also recall that preliminary Monte Carlo studies
     based on the lattice action (\ref{eLL}) support a nontrivial
     behavior of the $\p_4^4$ model exhibiting a positive renormalized coupling constant \cite{stank}.
     Additional Monte Carlo studies are welcome!

\subsection{Comparison of affine and canonical quantization results}
  There already exist well-defined, conventional results for $\p^4_2$ and $\p^4_3$, and, it is true, we are proposing alternative quantizations for these models. Unlike the usual approach, there are no divergences in perturbative expansions in the new formulation after operator product expansions are introduced for the local operators, and models like $g_{0}\s\p^4_3+g'_{0}\s\p^8_3$---a sum of a superrenormalizable and a nonrenormalizable interaction---exhibit reasonable properties, such as when either one of the coupling constants is sent to zero and then becomes positive again, the prior results apply. For $\p^4_4$, traditional methods lead to  a renormalizable theory, but nonperturbative methods support a trivial (free) behavior. On the other hand, our procedures are expected to be divergence free and nontrivial \cite{stank}. The cases $\p^4_n$, $n\ge5$, are conventionally nonrenormalizable and require an infinite number of distinct counterterms. In contrast, our formulation has been designed to have a divergence-free perturbation about the pseudofree theory and to yield a nontrivial result.
  %The success of the electroweak model leads to suggestions that nonrenormalizable models are effective %field theories,  good for low energy questions with a few perturbative corrections.

   An important key in constructing self-consistent solutions for nonrenormalizable models is the realization that the zero-coupling limit of interacting solutions is {\it different} from the usual free theory, e.g., as follows from (\ref{e333}); this distinction already holds for the classical theory and thus for the quantum theory. (Indeed, such behavior already arises for a {\it single particle} with the classical action
  $A=\tint\{\half[\s{\dot x}(t)^2-x(t)^2]-g\s x(t)^{-4}\s\}\,dt$!)
  Accepting that possibility opens the door for alternative counterterms that favor operator product expansions over normal ordering, and that can lead to divergence-free formulations.

  The inappropriateness of a perturbation analysis about the free model for the nonrenormalizable cases that follows from a difference
  between the pseudofree and free theory for such cases also eliminates the relevance of Landau poles for such
  problems. Those who believe the renormalization group is relevant are asked to consider the lattice action (\ref{eLL}) with its specific form and the special dependence of its coefficients on the lattice spacing $a$. The coefficients in the counterterm (\ref{WeF}) have been {\it designed} to lead to a nontrivial continuum limit. That special dependence on the lattice spacing would be hard to maintain throughout when taking the continuum limit via a renormalization group analysis.

\section{Summary} Conventional quantization of a classical system involves promoting suitable canonical coordinates to Hermitian operators and choosing a Hamiltonian operator for the Schr\"odinger equation of motion that follows the classical Hamiltonian up to possible corrections of order $\hbar$. This procedure works well for many systems but, alas, not for all systems. The triviality of $\phi^4_n$, $n>4$ (and perhaps $n=4$ as well) models is one example of a failure---or an unsatisfactory result, if you prefer---of conventional quantization. This has prompted a reexamination of the process of quantization, which leads to the enhanced quantization procedures (Sec.~1)
that establish a classical/quantum connection entirely within the quantum realm with $\hbar>0$ throughout.
For many systems, the alternative procedures lead to the very same results that arise using the original quantization procedures, thus reconfirming these satisfactory examples. However, the new procedures also allow for affine quantization techniques to replace canonical techniques, and this allows for satisfactory outcomes for certain models (Sec.~2, Sec.~4, and Sec.~6). The weak correspondence principle, which is the new relation between the quantum and classical Hamiltonians---indeed, it involves the enhanced classical Hamiltonian since $\hbar>0$---permits reducible quantum kinematical operators and thereby extends the range of satisfactory outcomes of quantization of given classical systems (Sec.~3). And last but not least, the beneficial results offered to the quantization of classical systems by means of an affine approach extend to gravity, and offer a conservative approach to that very difficult problem which already shows considerable promise (Sec.~5).

Finally, it would seem that the several examples discussed in this article are but a few of the models that can profit by reanalyzing their classical/quantum connection.

{\it Reference added after acceptance for publication:}  A very recently published book \cite{recent} builds on this article offering additional developments as well as several new examples of the power of enhanced quantization.

\subsection*{Acknowledgements}
It ia a pleasure to thank Prof.~E. Deumens and J.~Stankowitz for obtaining strong, preliminary, Monte Carlo evidence \cite{stank}
that the $\phi^4_4$ model augmented with the special counterterm (\ref{WeF}) has a positive, nonvanishing, renormalized coupling constant in the continuum limit.

The author is grateful to the Perimeter Institute, Waterloo, Canada, for its hospitality during the month of
May 2012, during which time this article was partially prepared. Research at the Perimeter Institute is supported by the Government of Canada through Industry Canada and by the Province of Ontario
through the Ministry of Research and Innovation.

Prof.~M.~Karasev, and the other organizers of the conference ``Interaction of Mathematics and Physics:
New Perspectives'', held in Moscow, Russia, August 22--30, 2012, are thanked for their support that
enabled the author's participation.

  \end{document}